\documentclass[aps,prd,amsmath,amssymb,amsfonts,floats,floatfix,superscriptaddress,nofootinbib,notitlepage,twocolumn]{revtex4-1}

\usepackage{hyperref}
\usepackage{epsfig}
\usepackage{epstopdf}
\usepackage{graphicx}
\usepackage{graphics}
\usepackage[retainorgcmds]{IEEEtrantools}
\usepackage{color}
\usepackage{xspace} 
\usepackage{multirow}

\allowdisplaybreaks[1]

\newcommand{\phiR}{\tilde{\Phi}^{\rm R}}
\newcommand{\tildephiS}{\tilde{\Phi}^{\rm S}}
\newcommand{\phiret}{\Phi^{\rm{ret}}}
\newcommand{\scriP}{\mathcal{J}^+}

\begin{document}
\title{Scalar self-force for eccentric orbits around a Schwarzschild black hole}

\author{Ian Vega}
\affiliation{SISSA - International School for Advanced Studies, Via Bonomea 265, 34136 Trieste, Italy and
INFN, Sezione di Trieste, Italy}
\affiliation{Department of Physics, University of Guelph, Guelph, Ontario, N1G 2W1, Canada}

\author{Barry Wardell}
\affiliation{School of Mathematical Sciences and Complex \& Adaptive Systems Laboratory, University College Dublin, Belfield, Dublin 4 Ireland}
\affiliation{Max-Planck-Institut f\"{u}r Gravitationsphysik, Albert-Einstein-Institut, 14476 Potsdam, Germany}

\author{Peter Diener}
\affiliation{Center for Computation \& Technology, Louisiana State University, Baton Rouge, LA 70803, U.S.A.}
\affiliation{Department of Physics \& Astronomy, Louisiana State University, Baton Rouge, LA 70803, U.S.A.}

\author{Samuel Cupp}
\affiliation{Department of Physics \& Astronomy, Austin Peay State University, Clarksville, TN 37044, U.S.A.}

\author{Roland Haas}
\affiliation{Theoretical Astrophysics 350-17, California Institute of Technology, Pasadena, California 91125, USA}

\begin{abstract}
We revisit the problem of computing the self-force on a scalar charge moving along an eccentric
geodesic orbit around a Schwarzschild black hole. This work extends previous scalar self-force
calculations for circular orbits, which were based on a regular ``effective" point-particle source
and a full 3D evolution code. We find good agreement between our results and previous calculations
based on a (1+1) time-domain code. Finally, our data visualization is unconventional: we plot the
self-force through full radial cycles to create ``self-force loops", which reveal many interesting
features that are less apparent in standard presentations of eccentric-orbit self-force data.
\end{abstract}
\maketitle

\section{Introduction}

Gravitational waves from highly relativistic systems such as compact object binaries are of
significant interest in astrophysics and fundamental physics. For astrophysics, gravitational waves
will eventually complement traditional observations based on electromagnetic waves, by allowing us
to peer through otherwise opaque regions of the cosmos \cite{Sathyaprakash:2009xs}. And for fundamental physics,
gravitational wave observations can serve as useful tools for probing strong-gravity phenomena,
supplementing the existing suite of weak-field, cosmological, and purely theoretical constraints on
alternative theories of gravity \cite{AmaroSeoane:2010zy}.

One very promising class of highly relativistic systems are binaries consisting of a massive black
hole (say of mass $m_1$) and a solar-mass compact object (of mass $m_2$), where $m_1 \gg m_2$.
These are known as EMRIs \cite{AmaroSeoane:2010zy,AmaroSeoane:2012km} --- short for \emph{extreme-mass-ratio inspirals} --- 
because of their general inspiraling behavior and the very small ratio ($q :=
m_2/m_1 \ll 1$) between the constituent masses. The existence of this small ratio makes it sensible
to adopt a perturbative strategy, whereby one considers the internal dynamics of the compact object
to be largely irrelevant to its bulk motion around the much heavier black hole. The small compact
object is thus seen as an inspiraling point mass that perturbs the spacetime of the black hole. In
the test-particle limit (or, equivalently, zeroth order in the mass ratio), the motion of the
particle is simply geodesic in the background spacetime, and for this case the technology for computing gravitational waves has been available since the 1970s \cite{davis-etal:71,detweiler:78}. This test-particle model, however, would be suboptimal for data
analysis purposes. Matched filtering, the standard method by which a weak gravitational wave signal
is extracted from a noisy data stream, requires that the phase of theoretical model waveforms
accurately matches that of the true signal throughout the detector sensitivity band. Otherwise, the
signal-to-noise ratio computed from a convolution of the template and the data can be significantly
diminished, causing one to completely miss a gravitational wave signal even if it really was present in
the data stream. It can happen that matched filtering with an inaccurate template still correctly infers the presence of a true
signal, but it does so at the price of associating the detected gravitational wave to wrong parameters for its astrophysical source. In either case, it is clear that errors in the waveform template
seriously undercut the practicability and utility of future gravitational wave observations.

\begin{figure}[h]
\begin{center}
\includegraphics[width=8.5cm]{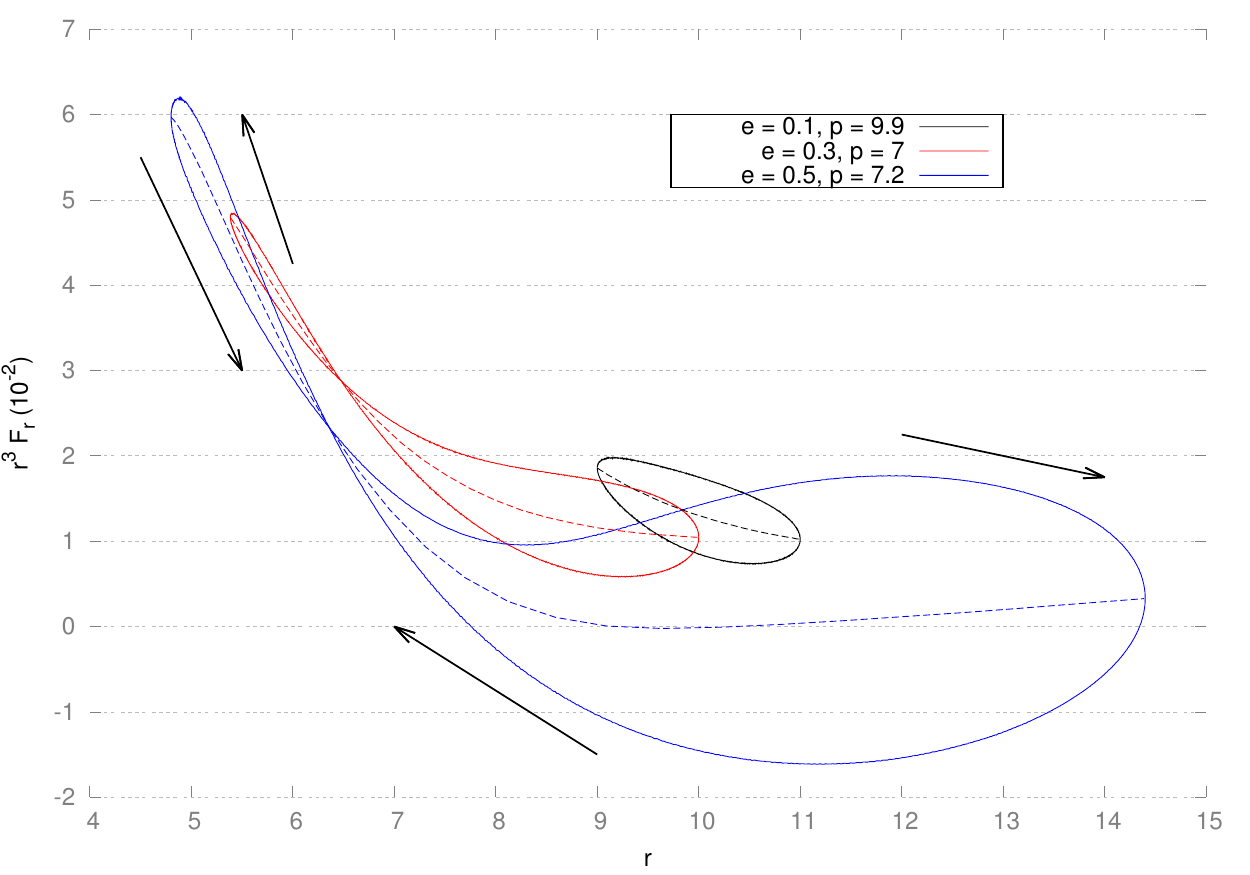}
\caption{Radial component of the self-force through one radial cycle. Solid lines
indicate the full self-force and dashed lines indicate the conservative-only piece. Eccentric orbits
that enter the strong-field region can experience a radial self-force which is stronger as the particle
moves inward in $r$ than as it moves outward; this is in contrast to the $t$ and
$\phi$ components (and to weak-field limits), where the outward motion always experiences a stronger
(or equal) self-force.}
\label{fig:frloop}
\end{center}
\end{figure}

With respect to point-mass models of EMRIs, this implies that simulations must include the influence
of the field (i.e., metric perturbation) generated by the point mass on its own motion. The modern
incarnation of the self-force problem is motivated principally by this need to incorporate as many
post-geodesic corrections as necessary to the motion of a point mass for a reasonably accurate
model waveform to be computed. This task is nontrivial in at least two respects: (1) the generated
field happens to be singular at the location of the point mass and is thus difficult to compute
(even numerically), and (2), owing to questions of gauge, inferring observable self-force effects
from the perturbation is conceptually challenging.

This paper focuses on the first of these difficulties, by further extending a method for
calculating self-forces first proposed in \cite{Barack:2007jh,Vega:2007mc}. The idea of this
approach is simple: to replace the traditional delta-function representation of a point source by
an appropriate regular \emph{effective source}, and thereby to deal only with fields that are
regular throughout the physical domain with no need for regularization. When it is implemented with a (3+1) evolution
code, such as those used in numerical relativity, the effective source approach is a powerful
strategy for simulating the self-consistent dynamics of particles and their fields
\cite{Diener:2011cc}. As a method for self-force calculation, this was previously demonstrated for
a scalar charged particle in circular orbits around the Schwarzschild geometry \cite{Vega:2009qb}.
The extension to eccentric orbits, while conceptually straightforward, has proven to be technically
challenging, primarily because constructing the effective source has been difficult. This
construction was eventually achieved and is described in \cite{Wardell:2011gb}. The present
manuscript showcases the use of this new effective source for self-force calculations for a
scalar charged particle moving along an eccentric geodesic of the Schwarzschild spacetime (see Fig.~\ref{fig:frloop}). Its
central point is that the effective source approach can accommodate a much larger class of orbits than
has been previously shown.
The present work allows us to assess the performance and merits of the method, and we do so
primarily by benchmarking our results against very accurate mode-sum computations based on a (1+1)
time-domain code. As a side note, we emphasize that the results of this paper were crucial to the
self-consistent simulations described in \cite{Diener:2011cc}.

The rest of the paper is as follows. In Sec.~II, after a short review of eccentric geodesics in the
Schwarzschild geometry, we present self-force results for the orbits we have analyzed and explain
their general features. Our results are illustrated as ``self-force loops", which essentially display the self-force as a function of the cyclic radial coordinate. We find this to be quite useful in visualizing eccentric-orbit self-force data. We also present the energy and
angular momentum losses through the event horizon and future null infinity, which are related to the cumulative action (of parts) of the local self-force on the particle. Section III discusses our
general calculational approach, which centers on an effective point-particle source evolved on a
(3+1) numerical grid. In Sec.~IV, we discuss more specific aspects of our simulations. We also
assess convergence and the accuracy of our methods by comparing against results
computed using a (1+1) mode-sum regularization code~\cite{Haas:2007kz}. We conclude in Sec.~V.

Throughout this paper, we use units in which $G=c=1$ and adopt the sign conventions of
\cite{Misner:1974qy}. Roman letters $i$, $j$ and $ k$ are used for indices over spatial dimensions
only, while Greek letters $\alpha, \beta, \ldots$ are used for indices which run over all spacetime
dimensions. Our convention is that $x$ refers to the point where a field is evaluated and
$\bar{x}$ refers to an arbitrary point on the world line. In computing expansions, we use
$\epsilon$ as an expansion parameter to denote the fundamental scale of separation, so that
$x-\bar{x} \approx \mathcal{O}(\epsilon)$. Where tensors are to be evaluated at these points, we
decorate their indices appropriately using $\bar{~}$, e.g. $T^a$ and $T^{\bar{a}}$ refer to tensors
at $x$ and $\bar{x}$, respectively.

\section{Self-force on eccentric orbits of Schwarzschild spacetime}

\subsection{Geodesics in the Schwarzschild geometry}

A test particle traces a geodesic in spacetime\footnote{We present here the bare minimum required
to understand the notation we use. For a more detailed treatment of geodesics in Schwarzschild
spacetime, see
\cite{Cutler:1994pb}, \cite{Pound:2007th} or \cite{Barack:2010tm}, from which we borrow much of our
discussion.}. In the case of the Schwarzschild spacetime, 
\begin{equation}
ds^2 = -\left(1-\frac{2M}{r}\right)dt^2+\left(1-\frac{2M}{r}\right)^{-1}dr^2 + r^2 d\Omega^2,
\end{equation}
with, $d\Omega^2 = d\theta^2 + \sin^2\theta d\phi^2$,
the Killing symmetries give two constants of motion
\begin{align}
-\mathcal{E} &:= t^\alpha u_\alpha = u_t \\
\mathcal{L} &:= \phi^\alpha u_\alpha = u_\phi
\end{align}
which are the particle's specific energy and angular momentum. The equations describing a timelike geodesic can then be written as: 
\begin{equation}
\frac{dt_p}{d\tau}= \mathcal{E}\left(1-\frac{2M}{r_p}\right)^{-1}, \quad \frac{d\phi_p}{d\tau} = \frac{\mathcal{L}}{r_p^2} 
\end{equation}
\begin{equation}
\frac{dr_p}{d\tau} = \pm\left[\mathcal{E}^2 - U_{\rm eff}(\mathcal{L};r_p)\right]^{1/2}
\label{eq:radialgeod}
\end{equation}
where the effective potential, $U_{\rm eff}(\mathcal{L};r)$, is 
\begin{equation}
U_{\rm eff}(\mathcal{L};r) := \left(1-\frac{2M}{r}\right) \left(1+\frac{\mathcal{L}^2}{r^2}\right)
\end{equation}
Here, we assume equatorial motion, $\theta_p=\pi/2$, which amounts to no loss in generality in the
Schwarzschild spacetime.

Bound orbits exist when $\mathcal{L}^2> 12M^2$. These orbits are uniquely specified by their inner
and outer radial turning points, or periastron ($r_{\rm min}$) and apastron ($r_{\rm max}$),
respectively. One convenient parametrization of these bound orbits makes use of the dimensionless
parameters $p$ and $e$, which are defined as
\begin{align}
p = \frac{2r_{\rm min}r_{\rm max}}{M(r_{\rm min}+r_{\rm max})}, \quad e = \frac{r_{\rm max}-r_{\rm min}}{r_{\rm max}+r_{\rm min}},
\end{align}
and correspond to the semilatus rectum and eccentricity of the (quasi-elliptical) orbit in the
weak-field regime. Intuitively, $p$ gives a sense of the size of the orbit, while $e$ has to do
with the orbit's shape. In this parametrization, the conserved quantities $\mathcal{E}$ and
$\mathcal{L}$ are given by
\begin{align}
\mathcal{E}^2 &= \frac{(p-2-2e)(p-2+2e)}{p(p-3-e^2)},\nonumber \\
\mathcal{L}^2 &= \frac{p^2 M^2}{p-3-e^2}.
\end{align}
Bound geodesics have $0 \leq e < 1$ and $p > 6 + 2e$. Points along the separatrix
$p = 6 + 2e$ (in which case the maximum of the effective potential is equal to $\mathcal{E}^2$)
represent marginally unstable orbits. Stable
circular orbits are those with $e = 0$ and $p \geq 6$, for which $\mathcal{E}^2$ equals the minimum
of the effective potential. The point $(p,e) = (6,0)$ in the $e$-$p$ plane, where the separatrix
intersects the $e = 0$ axis, is referred to as the innermost stable circular orbit (ISCO).

For this paper, the crucial property to note is that the fundamental periodicity for bound geodesics
 in Schwarzschild spacetime is set by the radial motion. Due to orbital precession, the system
(``particle" + ``field") is not periodic in $\phi$, but it nevertheless returns to an identical state with every full radial cycle. As such, all the
essential information concerning a radiating charge in a fixed eccentric orbit can be obtained from
one radial cycle; information from other cycles is redundant. In particular, this applies to the
self-force acting on this charge as well.

\subsection{Self-force}

By carrying a charge, the particle ceases to be a test body. The particle's charge gives rise to a
scalar field which interacts with the particle. Its path therefore deviates away from a geodesic due
to the action of the scalar self-force \cite{Quinn:2000wa}:
\begin{equation}
F_\alpha = q^2 (g_{\alpha\beta}+u_\alpha u_\beta)\left(\frac{1}{3}\dot{a}^\beta + \frac{1}{6}R^\beta{}_\gamma u^\gamma\right) + q \Phi_\alpha^{\rm tail}
\end{equation}
where 
\begin{equation}
\Phi_\alpha^{\rm tail} = q \int^{\tau-}_{-\infty} \nabla_\alpha G(z(\tau),z(\tau')) d\tau'
\end{equation}
is the nonlocal tail field and $G$ is the retarded Green function. The task at hand then lies in
calculating both the field and trajectory of the charged particle self-consistently. This is
directly analogous to the outstanding problem (mentioned in the Introduction) of computing the
self-forced orbit of a point mass and its corresponding gravitational waveforms.

In this paper (and several others \cite{Haas:2007kz,Haas:2011np,Canizares:2009ay,Canizares:2010yx,Canizares:2011kw,Warburton:2010eq,Warburton:2011hp}),
the physical picture is simpler and slightly different. Instead of computing the self-force and
trajectory consistently, we imagine keeping the particle on a fixed geodesic and ask what
\emph{external} force is necessary to keep the particle on the same orbit. To second order in $q$, the answer is what we present in this manuscript: a \emph{geodesic-based} self-force. We completely ignore the gravitational sector of this problem and argue that our results are valid in the regime for which $q \gg m$, where $m$ is the rest mass of the charged particle. There is also a metric perturbation induced by the stress-energy of the charge, but because the background is a vacuum spacetime, this metric perturbation is $O(q^2)$, which gives a smaller scalar self-force correction of $O(q^3)$. This is in contrast to the situation described in \cite{Zimmerman:2012zu}. 

While this simplification is made out of practical considerations, it is worth pointing out that
there are circumstances in which the geodesic self-force might be expected to very accurately
approximate the true self-force. When $q \ll M$, the deviation of the motion away from a geodesic
becomes so slow that the geodesic self-force becomes a good surrogate for the true self-force
\cite{Warburton:2011fk}. The extent to which this is true is a matter that demands further scrutiny.
Moreover, the geodesic self-force already displays much of the interesting and unintuitive features
of the true self-force, so it is useful for elucidating self-force physics, irrespective of
gravitational wave astronomy. And finally, because computing geodesic-based self-forces is in itself
a delicate numerical problem, it has proven to be an extremely useful benchmark for testing codes
and calculational methods. Indeed, this was the primary motivation for the present work.

\begin{figure}[htb!]
\begin{center}
\includegraphics[width=8.5cm]{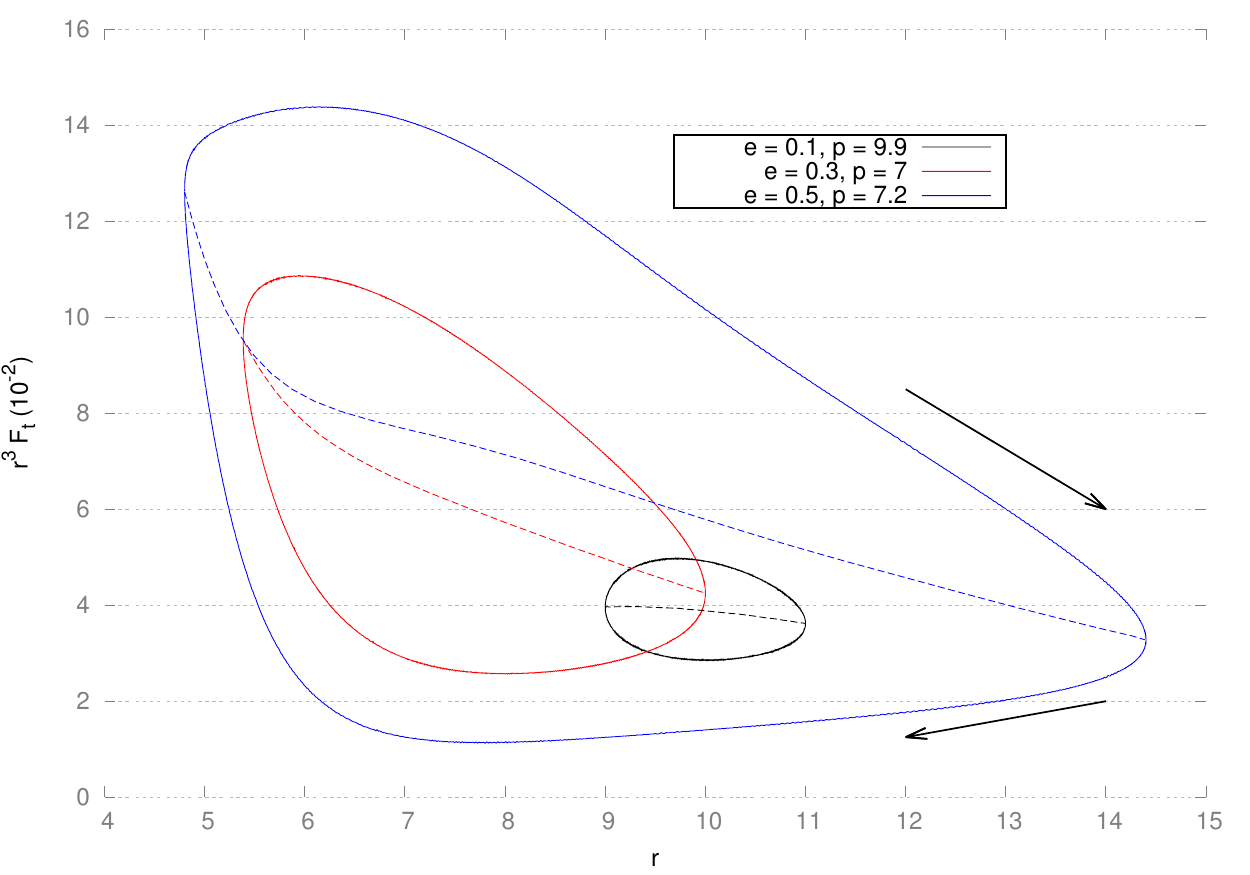}
\caption{Time component of the self-force through one full radial cycle. Solid lines
indicate the full self-force and dashed lines indicate the dissipative-only piece.}
\label{fig:ftloop}
\end{center}
\end{figure}

\begin{figure}[htb!]
\begin{center}
\includegraphics[width=8.5cm]{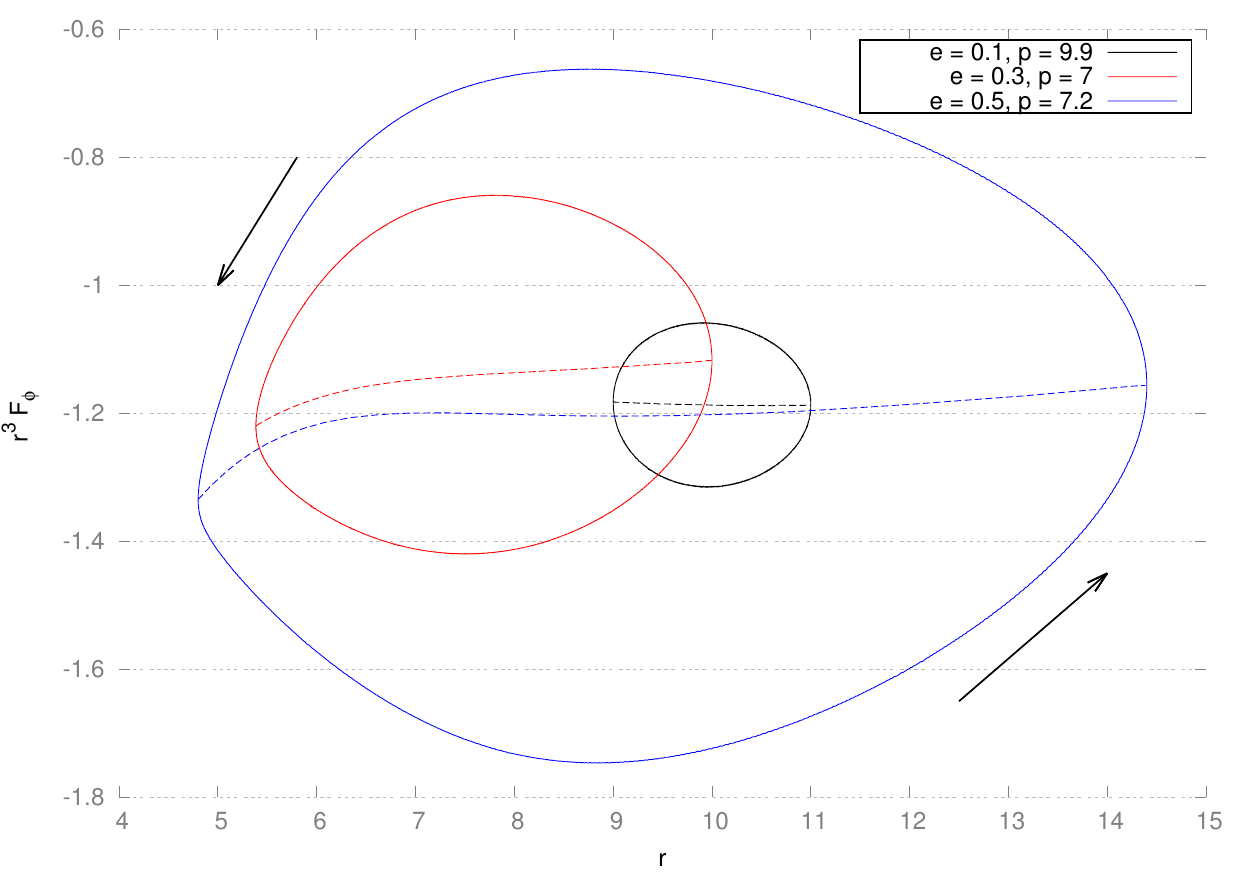}
\caption{Azimuthal component of the self-force through one full radial cycle. Solid lines
indicate the full self-force and dashed lines indicate the dissipative-only piece.}
\label{fig:fphiloop}
\end{center}
\end{figure}

Results from self-force calculations are typically presented as simple time series
\cite{Haas:2007kz,Haas:2011np,Canizares:2009ay,Canizares:2010yx,Canizares:2011kw,Warburton:2010eq,Warburton:2011hp}. We find it more illuminating,
instead, to plot the self-force as a function of the orbital radius. The self-force components are two-valued functions of the radial position of the particle, with each branch corresponding to either inward or outward radial motion
and therefore this creates closed loops like those shown in
Figs.~\ref{fig:frloop}, \ref{fig:ftloop}, and \ref{fig:fphiloop}. 
The arrows in these figures indicate the direction of the
particle's radial motion, and thus, also the direction of time evolution. Note that we have
factored out the gross $(1/r^{3})$-dependence of the self-force, which can be anticipated from
dimensional considerations.

From the figures, we see immediately that the self-force is generally different for inward and
outward motion. The self-force always weakens as the particle goes through apastron in each of our
three cases. (``Weaken" here means diminishes in strength or decreases in absolute value). This is
reversed at periastron, with the self-force strengthening \emph{after} the particle gets closest to
the black hole. A possible interpretation for this is that it is the retarded effect of scalar
field amplification occurring at periastron. But when the orbit gets sufficiently close to the black
hole (see Fig.~\ref{fig:frloop}), the peak of $F_r$ slightly \emph{precedes} periastron, and this
confuses the explanation. For these cases the loop twists before the particle reaches its closest
approach, so that there exists a crossover radial position where the radial component of the
self-force for outward and inward motion are equal. That this does not occur for our ``large-$p$,
low-$e$" case ($p=9.9, e=0.1$) suggests that it may be a signature of the strong-field regime, and
indeed, it is tempting to conjecture that this loop twisting is a general feature of orbits with
near-horizon periastra. Far enough from the black hole, the self-force is stronger for outward
motion than inward motion. Close to the black hole, this remains true for the $t$- and
$\phi$-components, but this behavior is reversed for the $r$-component.

More can be inferred from these loop figures. To appreciate this, we recall first that when
self-force effects on the orbital motion are small, these are often approximated by invoking
balance arguments for the conserved quantities and relying on averaged flux integrals to provide
the rates of change for the orbital parameters \cite{Hughes:2005qb,Drasco:2005is}. In this
\emph{adiabatic} approximation, the ``constants of motion" slowly change, and the particle
trajectory is replaced by a sequence of geodesics. Unfortunately, this scheme only picks up
\emph{dissipative} effects to the orbit, whereas the self-force affects the trajectory in ways that
cannot be associated with any balance law \cite{Pound:2005fs}. For this reason, extracting the
\emph{conservative} part of the self-force is then often\footnote{In fully self-consistent
simulations \cite{Diener:2011cc}, the split between dissipative and conservative pieces is
ambiguous. This decomposition is only really well defined for geodesic-based self-forces.} critical
in self-force calculations, if only to assess its importance.

Conservative and dissipative components of the self-force are defined to be those that are
symmetric and antisymmetric under the exchange ``retarded" $\leftrightarrow$ ``advanced"
\cite{Hinderer:2008dm,Barack:2010tm}, or equivalently, are are of even and odd parity with respect
to time reversal:
\begin{align}
F_{\alpha}^{\rm cons} &:= \frac{1}{2}(F_{\alpha}^{\rm ret}+F_{\alpha}^{\rm adv})\\
F_{\alpha}^{\rm diss} &:= \frac{1}{2}(F_{\alpha}^{\rm ret}-F_{\alpha}^{\rm adv})
\end{align}
where $F_{\alpha}^{\rm ret/adv}$ is the force resulting from retarded and advanced fields:
$F_{\alpha}^{\rm ret/adv}:= F_{\alpha}[\Phi^{\rm ret/adv}]$.

Taking $\tau_o$ to be proper time at either periastron or apastron, then in Schwarzschild
coordinates the retarded and advanced fields are related
\cite{Mino:2003yg,Hinderer:2008dm,Barack:2009ux,Barack:2010tm} according to
\begin{equation}
F_\alpha^{\rm adv}(\tau_o+\Delta \tau) = \epsilon_{(\alpha)}F_\alpha^{\rm ret}(\tau_o-\Delta \tau)
\label{eq:retadv}
\end{equation}
where $\epsilon_{(\alpha)} := (-1,1,1,-1)$. This allows us to write
\begin{align}
F_t^{\rm diss}(\tau_o+\Delta\tau) &= \frac{1}{2}\left[F_t^{\rm ret}(\tau_o+\Delta\tau)+F_t^{\rm ret}(\tau_o-\Delta\tau)\right] \label{eq:Ftdiss}\\
F_\phi^{\rm diss}(\tau_o+\Delta \tau) &= \frac{1}{2}\left[F_\phi^{\rm ret}(\tau_o+\Delta \tau)+F_\phi^{\rm ret}(\tau_o-\Delta\tau)\right] \label{eq:Fphidiss}
\end{align}
and
\begin{align}
F_r^{\rm cons}(\tau_o+\Delta\tau) &= \frac{1}{2}\left[F_r^{\rm ret}(\tau_o+\Delta \tau)+F_r^{\rm ret}(\tau_o-\Delta\tau)\right]. \label{eq:Frcons}
\end{align}
(These formulas are to be understood as having already been correctly regularized. The quantity on the right-hand side is, strictly speaking, the regularized self-force, $F_\alpha := \nabla_\alpha
\Phi^{\rm R}$. This is explained in Sec. \ref{sec:sfbasics}).

Now, since $dr_p/d\tau$
is purely a function of $r_p$, we can easily verify that $r_p(\tau_o+\Delta\tau) = r_p(\tau_o-\Delta\tau)$. Equations (\ref{eq:Ftdiss})-(\ref{eq:Frcons}) then mean that the simple averages of the top and bottom parts of the loops give the dissipative parts of
$F_t$ and $F_\phi$, and the conservative part of $F_r$. This average of the inward and outward
self-force components at each given value of $r$ is indicated by a dashed curve within each loop.
Correspondingly, the complement (i.e. difference between the dashed curve and the loop) gives the
dissipative part of $F_r$ and the conservative part of $F_t$ and $F_\phi$. Since these are
differences of the loop from its average, at any given $r$, the two differences should be equal in
magnitude but opposite in sign. Upon integration over one radial cycle then, only the contribution
from the dashed curve remains; time-averaged effects to the orbit are the result of the
conservative part of $F_r$ and the dissipative parts of $F_t$ and $F_\phi$.

More explicitly, assuming a unit mass for the particle, the change in its energy and angular
momentum through one radial cycle is
\begin{align}
-\Delta \mathcal{E} = \Delta u_t &= 2 \int_{r_{\rm min}}^{r_{\rm max}} \frac{F_t^{\rm diss}}{u^r} dr \label{eq:elossP}\\
\Delta \mathcal{L} = \Delta u_\phi &= 2 \int_{r_{\rm min}}^{r_{\rm max}} \frac{F_\phi^{\rm diss}}{u^r} dr. \label{eq:llossP}
\end{align}
Here, an additional term compensating for the mass loss (due to the tangential component of the scalar self-force) has been omitted as it averages to zero over a radial cycle
\cite{Warburton:2011hp}.

Note in the figures that $F_t^{\rm diss}>0$ and $F_\phi^{\rm diss} <0$, which implies that $\Delta
\mathcal{E} <0$ and $\Delta \mathcal{L} <0$. We confirm in the next subsection that these balance the total energy and angular momentum loss through the event horizon and future null infinity in the coordinate-time interval it takes the particle to go from $r_{\rm min}$ to $r_{\rm max}$.

Because an overall factor of $1/r^3$ is pulled out from the self-force in these figures, care must be exercised in visually comparing magnitudes at different radial positions. Nevertheless, the twisting of the $F_r$ loop is unmistakable; it signifies a sign change in the dissipative part of $F_r$ as the particle
gets close to the black hole. Again, it is tempting to speculate
that this is a generic feature of the strong-field regime.

These observed features can be usefully contrasted with the scalar self-force in the weak-field regime
\cite{Pfenning:2000zf}, which for a minimally coupled scalar field reads
\begin{equation}
{\bf f} = \frac{1}{3}q^2 \frac{d{\bf g}}{dt},
\end{equation}
where ${\bf g} := -\nabla \Phi({\bf x}) = -M/r$. This evaluates to 
\begin{equation}
{\bf f} = \frac{q^2M}{r^3}\left(\frac{2}{3}\dot{r}\hat{\bf r} -\frac{1}{3}r\dot{\phi}\hat{\boldsymbol{\phi}}\right).
\end{equation}
For minimal coupling, the weak-field scalar self-force is entirely dissipative.

As expected, the qualitative behavior of this weak-field self-force is consistent with the dissipative parts of
the full self-force when the particle nears apastron (i.e. farthest from the black hole). The dependence on the $\dot{r}$-factor is such that the dissipative radial component switches sign according to the direction of the radial motion: it is positive for outward motion and negative for inward motion. The dissipative azimuthal component similarly depends on $\dot{\phi}$, but does not change sign because the particle always moves in the direction of increasing $\phi$. 

The overall sign change of the dissipative $r$-component at somewhere other than the turning
points of the radial motion represents a stark deviation of the strong-field regime from the
weak-field qualitative behavior. Similarly, another deviation in qualitative behavior comes in the
most eccentric case we study, where the conservative piece of the the radial component also changes
sign during the orbit.

\subsection{Fluxes}

An important code check in this work is to compare the energy and angular momentum losses
computed from the local self-force with the corresponding fluxes through $\mathcal{J}^+$ and
the event horizon. This essentially tests the whole computational
infrastructure from the effective source itself to the hyperboloidal slicing,
wave equation integration and flux extraction. 
This equivalence can be shown mathematically \cite{Warburton:2011hp}, and it affirms our intuition concerning the basic physics of our
problem: the energy and angular momentum pumped into the charged particle to keep it moving along a
fixed geodesic must be that which escapes as radiative fluxes. 

\begin{table}
\begin{tabular}{|c|c|cc|cc|}
\hline
\multirow{2}{*}{$p$} & \multirow{2}{*}{$e$} & \multicolumn{2}{c|}{$10^4 \langle\dot{\mathcal{E}}\rangle$} & \multicolumn{2}{c|}{$10^3 \langle\dot{\mathcal{L}}\rangle$}\\
& & Self-force & Flux & Self-force & Flux \\
\hline
$9.9$ & $0.1$ & $-0.32880$ & $-0.32887$ & $-1.01025$ & $-1.01020$\\
$7.0$ & $0.3$ & $-1.6716$ & $-1.6715$  & $-2.6256$ & $-2.6252$\\
$7.2$ & $0.5$ & $-1.9682$ & $-1.9678$ & $-2.5867$ & $-2.5863$\\
\hline
\end{tabular}
\caption{Comparison of energy and angular momentum fluxes computing from the local self-force and
from flux extraction on the horizon and at $\cal J^+$.}
\label{table:flux}
\end{table}
Equations (\ref{eq:elossP}) and (\ref{eq:llossP}) give the change in energy and angular momentum
due to the local self-force through one radial cycle. The average losses per unit
time is then easily computed as $\langle\dot{\mathcal{E}}\rangle :=\Delta \mathcal{E}/T$ and
$\langle\dot{\mathcal{L}}\rangle :=\Delta \mathcal{L}/T$, where $T$ is the Schwarzschild time
interval between periastron and apastron. The resulting quantities are reported in the `Self-force'
columns of Table \ref{table:flux}. These are compared with corresponding averaged fluxes through
the event horizon and future null infinity. In Kerr-Schild coordinates on the horizon and
``Cartesian" hyperboloidal coordinates at $\mathcal{J}^+$, the angular momentum fluxes are
\begin{equation}
    \left.\frac{dL}{dt}\right|_{\mathcal{H}} = -\frac{M^2}{\pi}\oint_{r=2M} \frac{\partial\Phi}{\partial t}\left(x\partial_y\Phi-y\partial_x\Phi\right) d\Omega.
\label{eq:fluxH}
\end{equation}
\begin{equation}
\left.\frac{dL}{d\tau}\right|_{\mathcal{J}^+} = -\frac{\rho^2_{\scriP}}{4\pi}\oint_{\rho=\rho_{\scriP}} \frac{\partial\hat{\Phi}}{\partial\tau}\left(\hat{x}\partial_{\hat{y}}\hat{\Phi}-\hat{y}\partial_{\hat{x}}\hat{\Phi}\right)\; d\Omega.
\end{equation}
For the energy fluxes, we have
\begin{equation}
    \left.\frac{dE}{dt}\right|_{\mathcal{H}} = -\frac{M^2}{\pi}\oint_{r=2M} \left(\frac{\partial\Phi}{\partial t}\right)^2 d\Omega.
\label{eq:EfluxH}
\end{equation}
\begin{equation}
\left.\frac{dE}{d\tau}\right|_{\mathcal{J}^+} = -\frac{\rho^2_{\scriP}}{4\pi}\oint_{\rho=\rho_{\scriP}} \left(\frac{\partial\hat{\Phi}}{\partial\tau}\right)^2\; d\Omega.
\end{equation}
Here, an overbar denotes quantities in the conformally rescaled, hyperboloidal slicing modification
of the Kerr-Schild spacetime used in our numerical code \cite{Vega:2011wf}.
Such hyperboloidal slicings was described in general in 
\cite{Zenginoglu:2007jw} and specialized to this particular case in
\cite{Zenginoglu:2009hd}.
Derivations for these flux expressions can be found in the Appendix, except for Eq. (\ref{eq:EfluxH}), which is already derived in \cite{Vega:2009qb}. 

Integrating these over one radial cycle -- which is independent of whether Schwarzschild, Kerr-Schild or hyperboloidal coordinates are used -- gives the
values in the `Flux' column of Table \ref{table:flux}. Quite notable is the level of agreement in
the calculated average quantities; they differ at most by $0.02\%$.

\section{Methods of calculation}

\subsection{Field equation and self-force}
\label{sec:sfbasics}

The main idea underlying the \emph{effective source approach} is to replace a delta-function
point-particle source with a regular source. Typically, the first step in a traditional self-force
calculation is to solve the wave equation,
\begin{equation}
    \Box \Phi^{\rm{ret}} = -4\pi q \int \delta^{(4)}(x-z(\tau))d\tau,
\end{equation}
for the retarded field sourced by a point-particle charge $q$ whose world line, $\gamma$, is described by
$z(\tau)$. This retarded field is singular along $\gamma$, and thus requires a regularization
procedure in order to extract the piece of the field responsible for the self-force. In the
effective source approach, we instead work with
\begin{equation}
    \Box \phiR = S(x,z(\tau)), 
    \label{eqn:phiR}
\end{equation}
where $S(x,z(\tau))$ is constructed to be regular along $\gamma$. This results in the field,
$\Phi^{\rm R}$, also being regular along $\gamma$. The crux of the method lies in constructing $S$
as follows:
\begin{equation}
    S := -4\pi q\int \delta^{(4)}(x-z(\tau)) d\tau - \Box \tilde{\Phi}^{\rm {S}},
    \label{eqn:S}
\end{equation}
where $\tilde{\Phi}^{\rm S}$ is a reasonably accurate approximation to the Detweiler-Whiting singular
field \cite{Detweiler:2002mi}, which has been shown to play no role in the dynamics of the scalar
charge (apart from renormalizing its mass). By construction, the Detweiler-Whiting singular field
satisfies
\begin{equation}
    \Box \tildephiS = -4\pi \int \delta^{(4)}(x-z(\tau)) d\tau + \Delta(x,z(\tau)), \,\, x \in \mathcal{N}(z),
\end{equation}
where, for some measure of distance, $\epsilon$, away from the world line $z(\tau)$, the residual
field $\Delta(x,z(\tau)) = O(\epsilon^n)$ as $\epsilon \rightarrow 0$. The construction is strictly
defined only when the field point $x$ is within the normal neighborhood of the world line,
$\mathcal{N}(z(\tau))$.

Note that, by definition, the d'Alembertian of the
singular field exactly cancels the delta function on the world line and so in practical terms the computation of the effective source amounts to computing the d'Alembertian of the singular
field at all other points. 

For the region outside $N(z)$, there are various options. One may choose to use $S = \Delta$ to
solve for $\phiR$ \emph{only} inside $N(z)$ (or some subregion of it, such as a narrow worldtube, for
example, in \cite{Barack:2007jh,Dolan:2010mt,Dolan:2011dx}) and then ``switch variables" outside
this region, so that one solves for a $\Phi^{\rm{ret}}$ satisfying the vacuum field equation
instead. Attention must then be given to enforcing matching conditions for $\phiR$ and
$\Phi^{\rm{ret}}$ at the boundary separating the computational domains.

Another option, which is the one adopted here, is to use
\begin{equation}
    S := -4\pi q\int \delta^{(4)}(x-z(\tau)) d\tau - \Box \left(W\tilde{\Phi}^{\rm {S}}\right) = \tilde{\Delta}(x,z(\tau)),
\end{equation}
where $\tilde{\Delta}(x,z(\tau)) = O(\epsilon^n)$ and where $W$ is
a smooth ``window" function such that $W(z) =1, (\nabla_\alpha W)|_{x=z} = 0$ and $W(x) = 0$ when
$x \notin \mathcal{N}(z)$. The first two conditions ensure that the window function does not affect
the value of the calculated self-force, while the last condition obviates the need for separate
computational domains, since one can now just safely use $S= \tilde{\Delta}$ even outside the normal
neighborhood, but at the cost of complicating the effective source.

Linearity of the field equation implies that, in solving \eqref{eqn:phiR} for some specified
$\gamma$, we get
\begin{equation}
    \phiR = \phiret - \tildephiS,
\end{equation}
and according to \cite{Detweiler:2002mi}, assuming there is no external scalar field, the acceleration of the particle is then simply
\begin{equation}
    ma^\alpha = q(g^{\alpha\beta}+u^\alpha u^\beta)\nabla_\alpha \phiR |_{x=z}.
    \label{eq:sf}
\end{equation}
Strictly speaking, the self-force captures all $O(q^2)$ interaction effects between the scalar charge and its field, whereas the equation above projects out only the piece that is orthogonal to the world line (i.e. it is the self-acceleration). In the scalar field case considered here, there may also be a component tangent to the world line, which
results in a change in the mass of the particle, according to \cite{Quinn:2000wa}. For ease of exposition, we discuss the full self-force from which the orthogonal and tangential components can readily be obtained.

\subsection{Effective source}
\label{sec:effective-source}

When numerically evolving Eq.~\eqref{eqn:phiR}, we require an explicit expression for
$S(x,z(\tau))$ written in the coordinates of the background spacetime. As can be seen from its
definition in Eq.~\eqref{eqn:S}, this only requires an explicit coordinate expression for the
Detweiler-Whiting singular field. Originally, such a coordinate expression was only available for a
scalar charge in a circular orbit on a Schwarzschild background spacetime, written in terms of
standard Schwarzschild coordinates \cite{Detweiler:2002gi}. More recently, Haas and Poisson
\cite{Haas:2006ne} derived a covariant expression valid for arbitrary coordinate choices.

Their strategy was to first develop a covariant expansion of the Detweiler-Whiting singular field,
and then to write coordinate expressions for the elements of the covariant expansions. From
\cite{Haas:2006ne}, and relying on the bitensor formalism described in \cite{Poisson:2011nh}, a
covariant expansion for the Detweiler-Whiting singular field reads
\begin{eqnarray}
\label{eqn:PhiScov}
&& \Phi^{\rm S}(x,\bar{x}) \approx q\Bigg\{\frac{1}{s} + \bigg[\frac{\bar{r}^2-s^2}{6 s^3} R_{u \sigma u \sigma}\bigg] \nonumber \\
&& \quad +  \frac{1}{24 s^3} \bigg[ \left(\bar{r}^2 - 3 s^2\right) \bar{r} R_{u \sigma u \sigma | u} -
\left(\bar{r}^2-s^2\right) R_{u \sigma u \sigma | \sigma} \bigg] \Bigg\},\nonumber \\
\end{eqnarray}
where we have neglected terms of $\mathcal{O}(\epsilon^3)$ and higher. Here, $\bar{x}$ is a point on
the world line connected to the field point $x$ by a unique spacelike geodesic,
$s^2 := (g^{\bar{\alpha} \bar{\beta}} + u^{\bar{\alpha}} u^{\bar{\beta}}) \sigma_{\bar{\alpha}} \sigma_{\bar{\beta}}$  (i.e.\
the projection of $\sigma_{\bar{a}}$ orthogonal to the world line), $\bar{r} := \sigma_{\bar{\alpha}} u^{\bar{\alpha}}$
(the projection along the world line) and $R_{u \sigma u \sigma | \sigma} := \nabla_{\bar{\epsilon}} R_{\bar{\alpha} \bar{\beta} \bar{\gamma}
\bar{\delta}} u^{\bar{\alpha}} \sigma^{\bar{\beta}} u^{\bar{\gamma}} \sigma^{\bar{\epsilon}}
\sigma^{\bar{\delta}}$. The inverse metric and four-velocity of the particle evaluated at $\bar{x}$
are denoted by $g^{\bar{\alpha}\bar{\beta}}$ and $u^{\bar\alpha}$, respectively. The key
expansion element here is the bitensor 
$\sigma_{\bar\alpha}(x,\bar{x}) := \nabla_{\bar\alpha}\sigma(x,\bar{x})$, where Synge's world function
$\sigma(x,\bar{x})$ is defined as half the squared geodesic distance between $x$ and $\bar{x}$:
\begin{equation}
    \sigma(x,\bar{x}) := \frac{1}{2}\int g_{\alpha\beta}\frac{dy^\alpha}{d\lambda}\frac{dy^\beta}{d\lambda} d\lambda,
\end{equation}
and $y(\lambda)$ is the unique spacelike geodesic that links $x$ and $\bar{x}$: 
$y(\lambda=0)=\bar{x}$, $y(\lambda=1) = x$. The quantity $\sigma_{\bar{\alpha}}(x,\bar{x})$ serves as a covariant measure of 
distance between $x$ and $\bar{x}$.

Combining \eqref{eqn:PhiScov} with a coordinate expansion of $\sigma_{\bar{\alpha}}$, we have a
complete coordinate expression for the Detweiler-Whiting singular field valid within a normal
neighborhood of the world line. Note that this is generic since $u^{\bar{\alpha}}$ is left
unspecified; the only assumptions we have made are that the spacetime is vacuum and asymptotically
flat, and that the world line is a geodesic of the background. In the present context,
we work with the Schwarzschild spacetime in the Kerr-Schild coordinates used by our
evolution code. To produce a global extension of our definition of the singular field, we
choose $x$ and $\bar{x}$ so that they have the same Kerr-Schild time coordinate. This gives
us an expression for the singular field of the form
\begin{equation}
\label{eq:PhiS-coord}
\tilde{\Phi}_{\rm S} = \frac{a_{(6)} + a_{(7)} + a_{(8)} + a_{(9)}}{(b_{(2)})^{7/2}},
\end{equation}
where we introduce the notation for a term of order $n$,
$a_{(n)} = a_{i_1 \cdots i_n} (t, r, \phi)\Delta x^{i_1} \cdots \Delta x^{i_n} $.
Finally, we further manipulate this expression, making it periodic in the $\phi$ direction
and multiplying by the spatial window function (introduced in the previous section)
which goes to $0$ away from the world line before any
coordinate singularities are encountered. The full details of this effective source construction
procedure are discussed in much more detail in a separate paper \cite{Wardell:2011gb}.

\subsection{Evolution code}

We numerically evolve the sourced scalar wave equation, Eq.~\eqref{eqn:phiR}, on a fixed
Schwarzschild background spacetime using a spherical, $6$-block computational domain with $8$-th
order spatial finite differencing and $4$th-order Runge-Kutta time integration. The code --- which
is based on components of the \textsc{Einstein Toolkit} \cite{Loffler:2011ay}, in particular the
\textsc{Cactus} framework \cite{Goodale:2002a,Cactuscode:web} and the \textsc{Carpet}
\cite{Schnetter:2003rb,CarpetCode:web} adaptive mesh-refinement driver --- is
described in more detail in~\cite{Schnetter:2006pg}; here we only summarize its key properties. We
use touching blocks, where the finite differencing operators on each block satisfy a summation-by-parts
property and where characteristic information is passed across the block boundaries using
penalty boundary conditions. Both the summation by parts operators and the penalty boundary
conditions are described in more detail in~\cite{Diener:2005tn}. The code has been extensively
tested, having been used to perform simulations of a scalar field interacting with a Kerr black
hole \cite{Dorband:2006gg} and to compute the self-force on a scalar charge in a circular geodesic
orbit around a Schwarzschild black hole \cite{Vega:2009qb}. Our primary modifications to the code
relative to the previous, circular orbits version were to replace the effective source with the one
described in Sec.~\ref{sec:effective-source} and to modify the coordinates of the background
spacetime such that they give a hyperboloidal slice of the Schwarzschild spacetime in the wave zone
with a smooth transition to a Kerr-Schild slice in the near-zone. We ensure that this near-zone
region entirely covers the region of support of the effective source.

We compute the particle orbit using the geodesic\footnote{The computed
self-force is \emph{not} used to drive the orbital motion, unlike the self-consistent calculation in \cite{Diener:2011cc}.} equations in Kerr-Schild coordinates (our slicing is such that the
orbit is always within the Kerr-Schild region of the spacetime). In doing so, we use the same
Runge-Kutta time integration routines with the same time step as for the scalar field evolution. We
compute the self-force by interpolating the derivatives of the field to the world-line position
using $4$th order Lagrange polynomial interpolation.

\section{Numerical checks}

\subsection{Summary of simulations}

\subsubsection{Numerical grid parameters}
All simulations were performed using a spherical, 6-block system with $60$, $80$ and $100$ angular
cells per block and corresponding radial resolutions of $0.1M, 0.075M$ and 
$0.06M$ for low, medium and high resolutions, respectively. 
We evolved with hyperboloidal coordinates of the form described in~\cite{Zenginoglu:2007jw,Zenginoglu:2009hd,Vega:2011wf}), with parameters such that the inner
boundary was inside the horizon at $r_{\rm in}=\{1.8M, 1.775M, 1.76M\}$ for the 
three different resolutions, the transition from Kerr-Schild to
hyperboloidal slicing happened in the region $25M > r > 85M$ and the outer boundary at
$r_{\rm out}=\{100M, 100.025M, 100.04M\}$ corresponded to $\cal J^+$. 
The choice of the slightly different values for $r_{\rm in}$ and $r_{\rm out}$
for the different resolutions was dictated by our need to have grid points
located precisely at the horizon ($r=2M$) for clean extraction of the horizon
fluxes.
In the transition region, we used the smooth
transition function
\begin{equation}
\label{eq:transition}
 f(x) =
 \dfrac{1}{2}  + \dfrac{1}{2}\tanh \left( \dfrac{s}{\pi} \left\{
 \dfrac{\tan^2\left[\dfrac{\pi}{2w}(x-x_0)\right]-q^2}{\tan\left[\dfrac{\pi}{2w}(x-x_0)\right]}\right\}\right)
\end{equation}
with $x=r$, $x_0=25M$, $w=60M$, $q=1$ and $s=2$. At both inner and outer boundaries the geometry ensured
that all characteristics left the computational domain so that there were no incoming modes and
therefore boundary conditions were unnecessary. We used the 8-4 diagonal norm summation by parts
finite differencing operators and added some compatible explicit Kreiss-Oliger dissipation to all evolved variables. We
set the scalar field and its derivatives to $0$ initially and evolved the system until the
transient ``junk radiation'' dissipated, typically over the timescale of one orbit. We verified
that this was the case by checking that the computed self-force was periodic with the same period
as the orbit.

\subsubsection{Orbital configurations}
We studied three different orbital configurations with eccentricity $e = \{0.1, 0.3, 0.5\}$
and semilatus rectum $p = \{9.9M, 7.0M, 7.2M\}$, respectively. In all cases we used the smooth
transition window function \eqref{eq:transition} to restrict the support of the effective source
to the vicinity of the world line. In the polar direction, we chose $x=\theta$,
$x_0=\tfrac{\pi}{2} \pm 0.1$, $w= \pm 1.2$, $q=1$ and $s=2.25$. In the region outside the orbit
(toward $\cal J^+$), we chose $x=r$, $x_0=\{16M, 16M, 15.4M\}$, $w=9M$, $q=1$ and
$s=2.2$, for $e=\{0.1, 0.3, 0.5\}$, respectively. In the region inside the orbit (toward the horizon), we found that it was not necessary
to use a window function at all. However, we did have to add back the 
singular part of the field before integrating the flux across the horizon. 
This particular set of parameters was chosen by experimentation ---
using too narrow a window function leads to steep gradients and large numerical error, while using
too wide a window function means that the effective source must be evaluated at a large number of
grid points, significantly impacting the run time of the code. It is worth noting, however, that
the extracted self-force is independent of the choice of window function parameters, as expected.

\subsection{Error analysis}
\subsubsection{Validation against (1+1) time-domain results}
\begin{figure}[t!]
\begin{center}
\includegraphics[width=8.5cm]{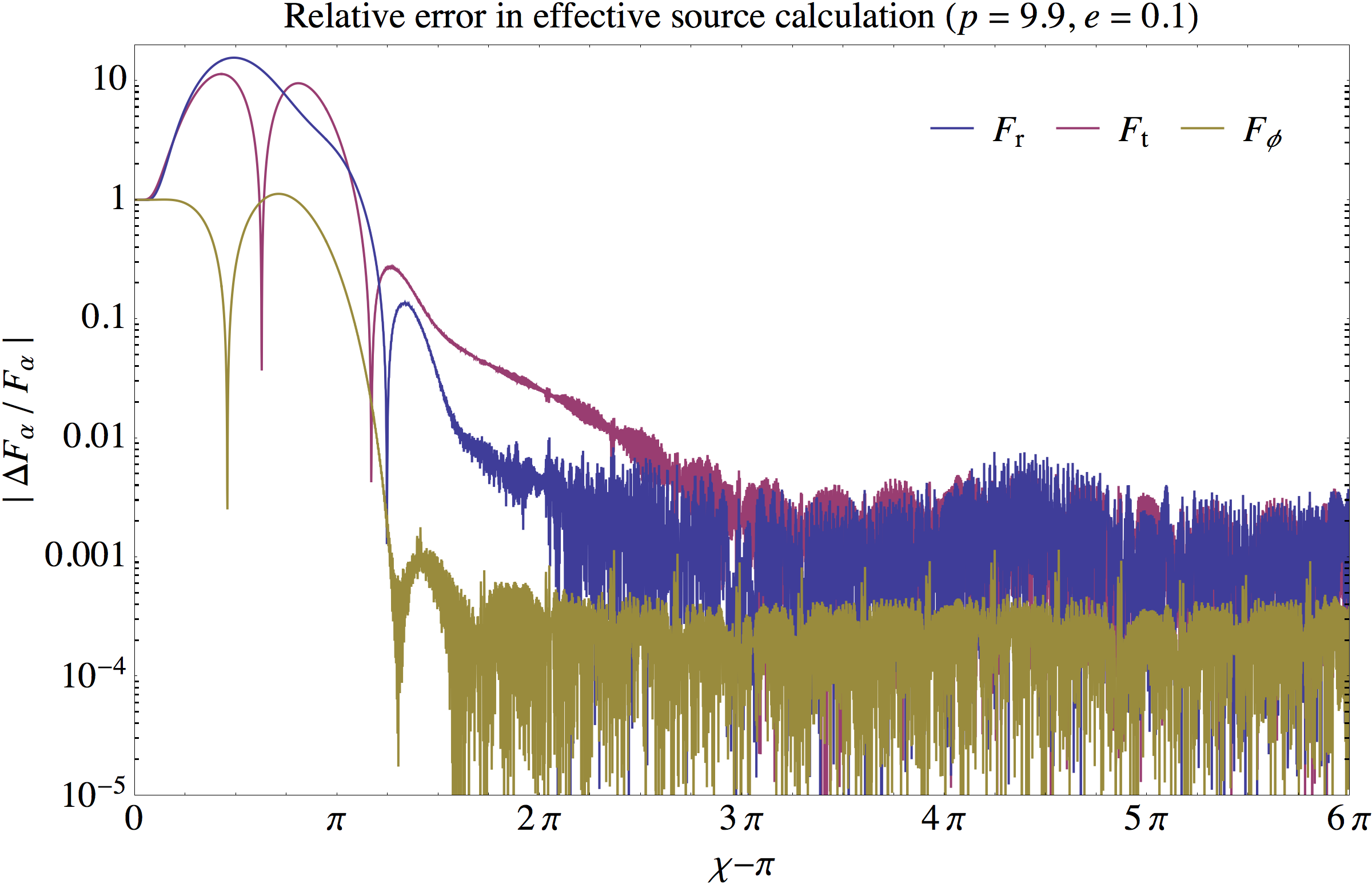}
\includegraphics[width=8.5cm]{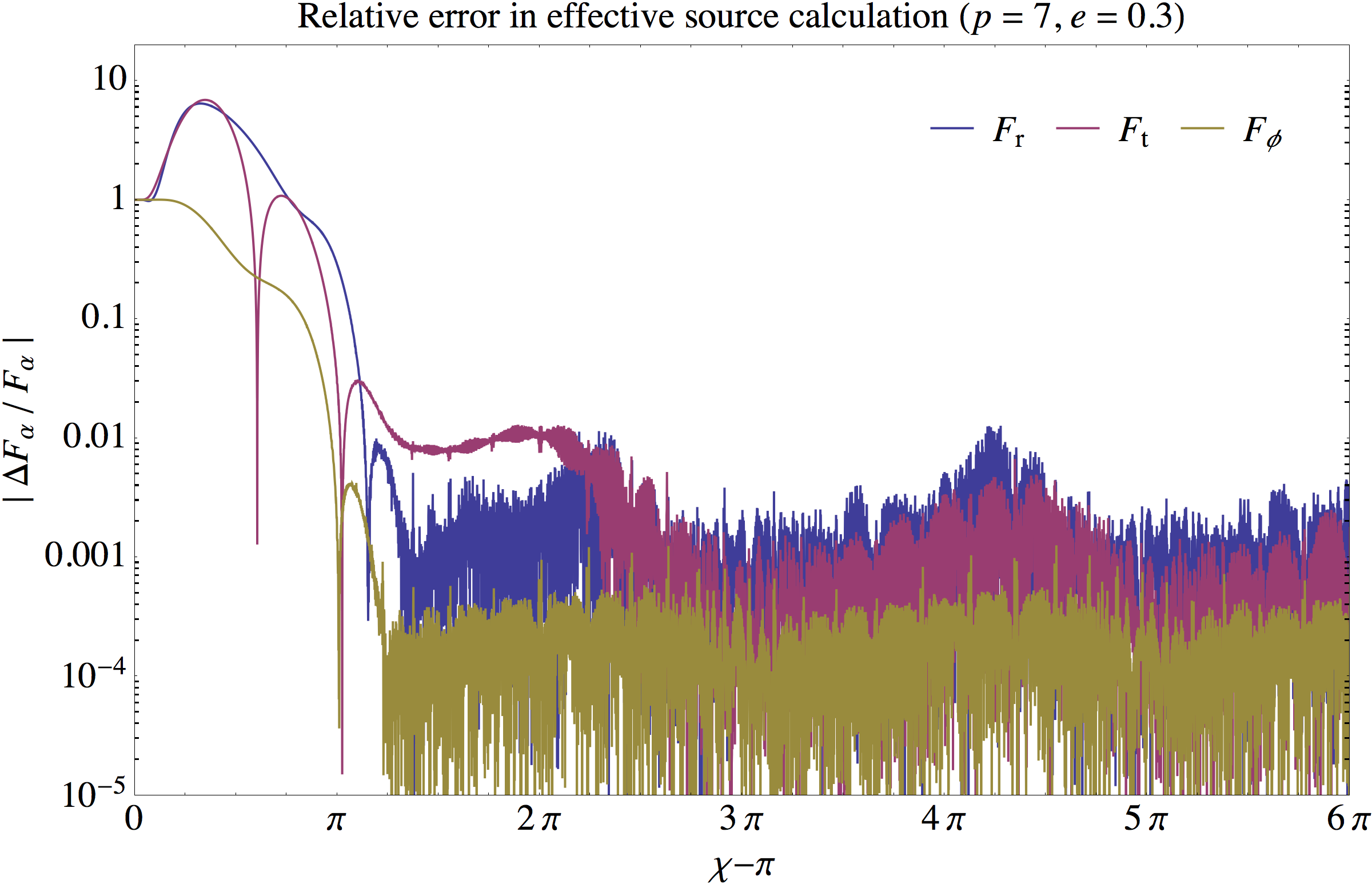}
\includegraphics[width=8.5cm]{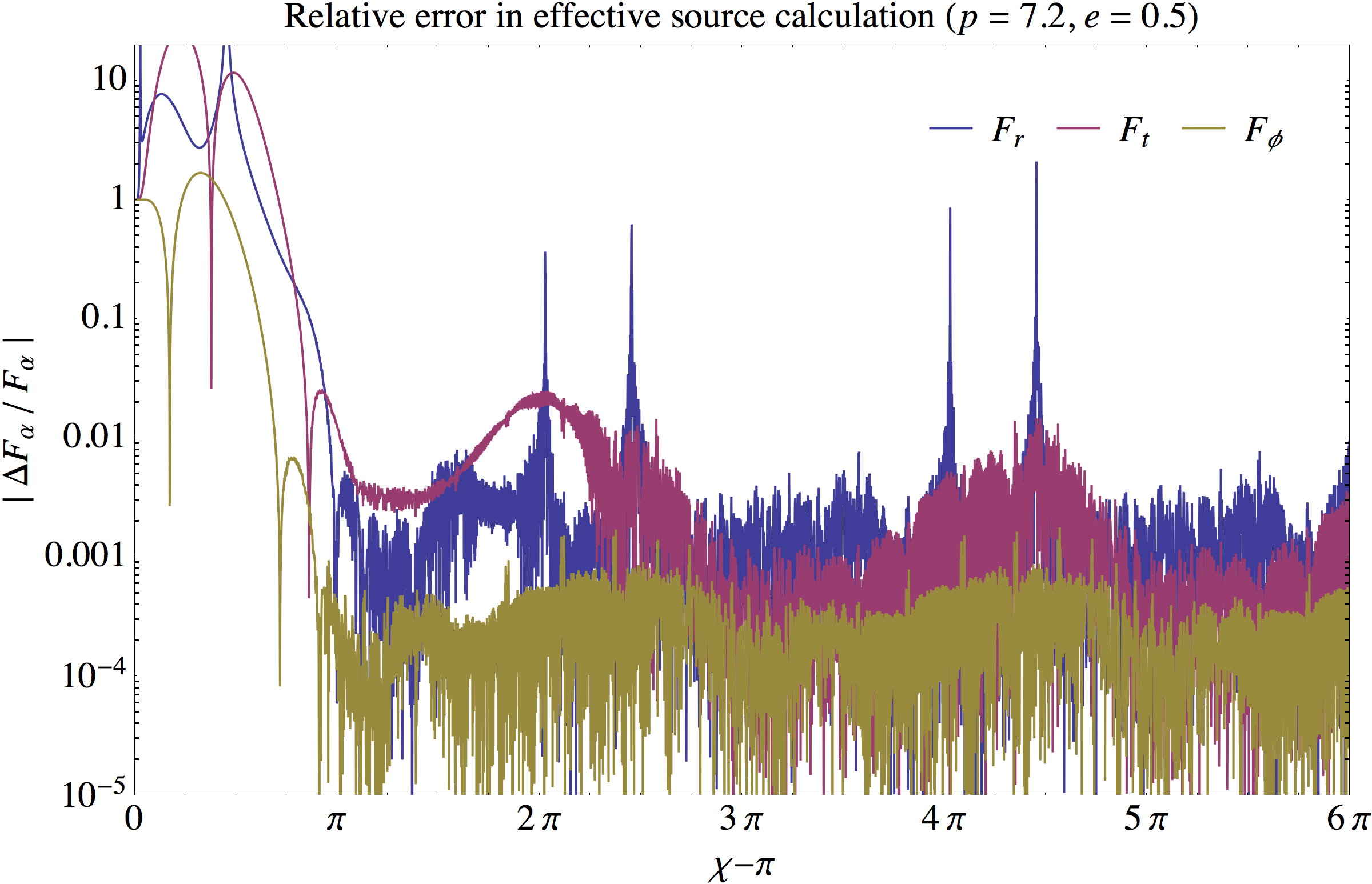}
\caption{Relative error in the self-force for the three orbital configurations considered.
Note that in the $e=0.5$ case the radial component passes through zero around
$\chi-\pi \approx 2 n \pi$ and $\chi-\pi \approx 2 n \pi + \tfrac{\pi}{2}$, for all integers $n$. As such we interpret the spikes in the
relative error at these points as merely an artifact of this zero-crossing.}
\label{fig:relative-error}
\end{center}
\end{figure}

For eccentric orbits, the three components of the self-force are independent of each other. (This
is in contrast to the circular orbit case, where the helical symmetry of the system relates the
$t$- and $\phi$-components). The plots in Fig.~\ref{fig:relative-error} show the relative error,
\begin{equation}
  |\Delta F_\alpha / F_\alpha| \equiv |1 - F_\alpha / F_\alpha^{\rm ref} |,
\end{equation}
for the highest resolution in each of the three self-force components for the three specific cases 
that were simulated. Reference values for the self-force were computed using the (1+1) time
domain code described in \cite{Haas:2007kz}.

We see that the initial burst of junk radiation (coming from inconsistent
initial data) contaminates the self-force for up to one orbit. After the junk radiation has
radiated away, the self-force settles down to within $1\%$ of the
reference value. The high-frequency oscillations in the error reflect the fact that the
low-order differentiability of the solution on the world line introduces a finite differencing error
which oscillates at the frequency with which the world line moves from one grid point to the next.
This could be improved by using a higher order approximation to the
singular field, thereby increasing the smoothness of the solution. This benefit
would, however, come at the cost of a substantially more complicated (and computationally costly)
effective source.

\subsubsection{Convergence}
Our evolution code has been shown to converge cleanly at the expected order when evolving
smooth initial data~\cite{Diener:2005tn}. The convergence order is determined both by 
the order of finite differencing in the interior region and at the inter-patch boundaries.
For example, for the 8-4 summation by parts operators used here, fifth order global
convergence is to be expected.

However, our choice of approximation to the singular field yields an effective source which
is only $C^0$ on the world line of the particle, and the evolved residual field is therefore
$C^2$ at the same point. Elsewhere, the solution is expected to be perfectly smooth.
Unsurprisingly, this lack of smoothness spoils any hope of clean high-order convergence of the
solution. It was shown in Appendix~A of \cite{Vega:2009qb} that for the wave equation in
1+1D, the errors are instead expected to converge at second order in the L2-norm for a
$C^0$ source. It is also shown that the error is of high frequency with the
frequency increasing with resolution. Thus, we cannot demonstrate pointwise convergence
for the evolved fields; instead we expect that the amplitude of any
noise generated near the world line will converge away at second order.

\begin{figure}[htb!]
\begin{center}
\includegraphics[width=8.5cm]{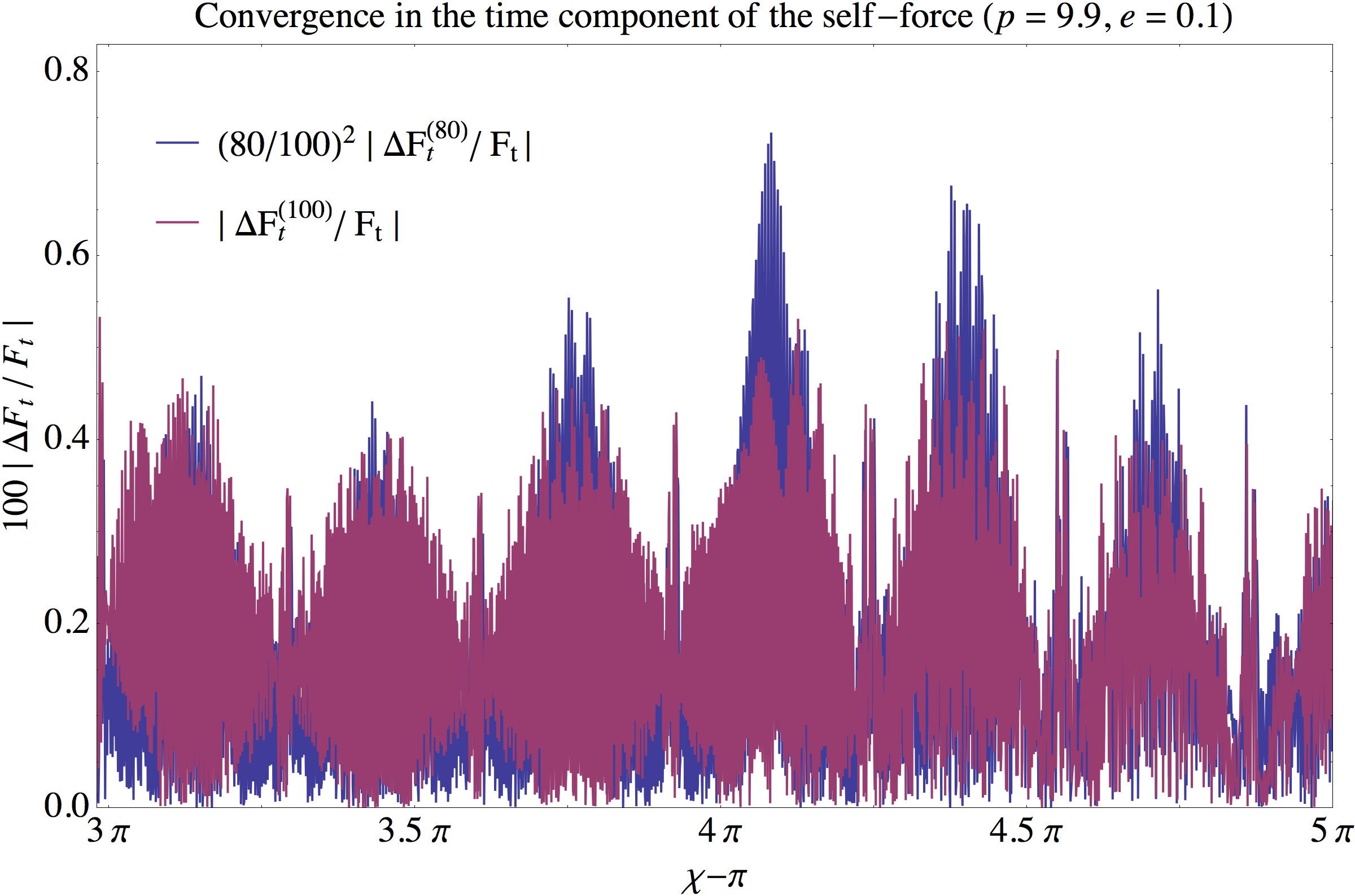}
\caption{Relative error in the $t$ component of the self-force for the $e=0.1$, $p=9.9$
case. When rescaled by the anticipated second-order convergence factor, the errors in the high
resolution simulation are comparable to those of the medium resolution.
}
\label{fig:convergence-ft}
\end{center}
\end{figure}
\begin{figure}[htb!]
\begin{center}
\includegraphics[width=8.5cm]{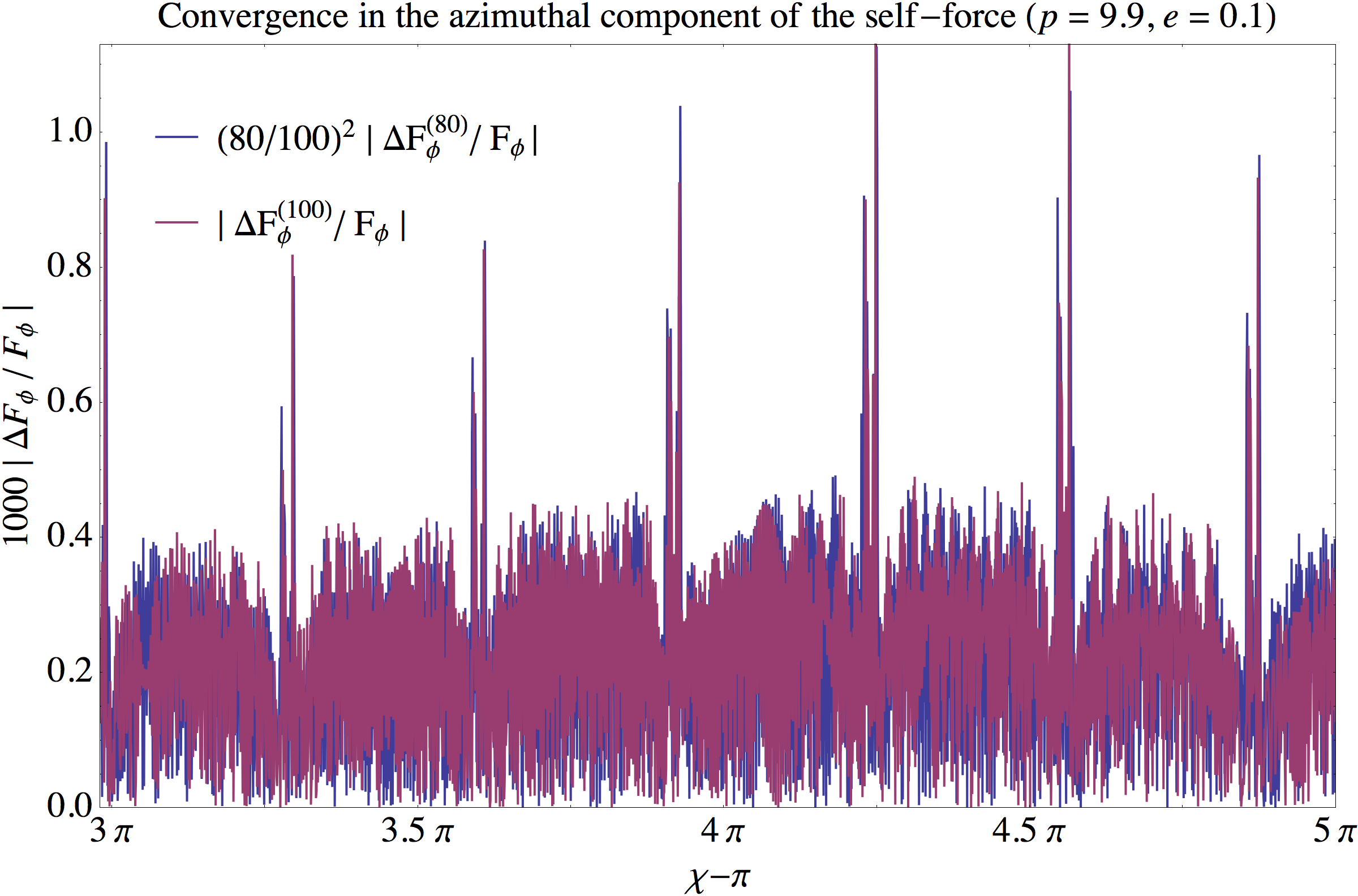}
\caption{Relative error in the $\phi$ component of the self-force for the $e=0.1$, $p=9.9$
case. When rescaled by the anticipated second-order convergence factor, the errors in the high
resolution simulation are comparable to those of the medium resolution.}
\label{fig:convergence-fphi}
\end{center}
\end{figure}
\begin{figure}[htb!]
\begin{center}
\includegraphics[width=8.5cm]{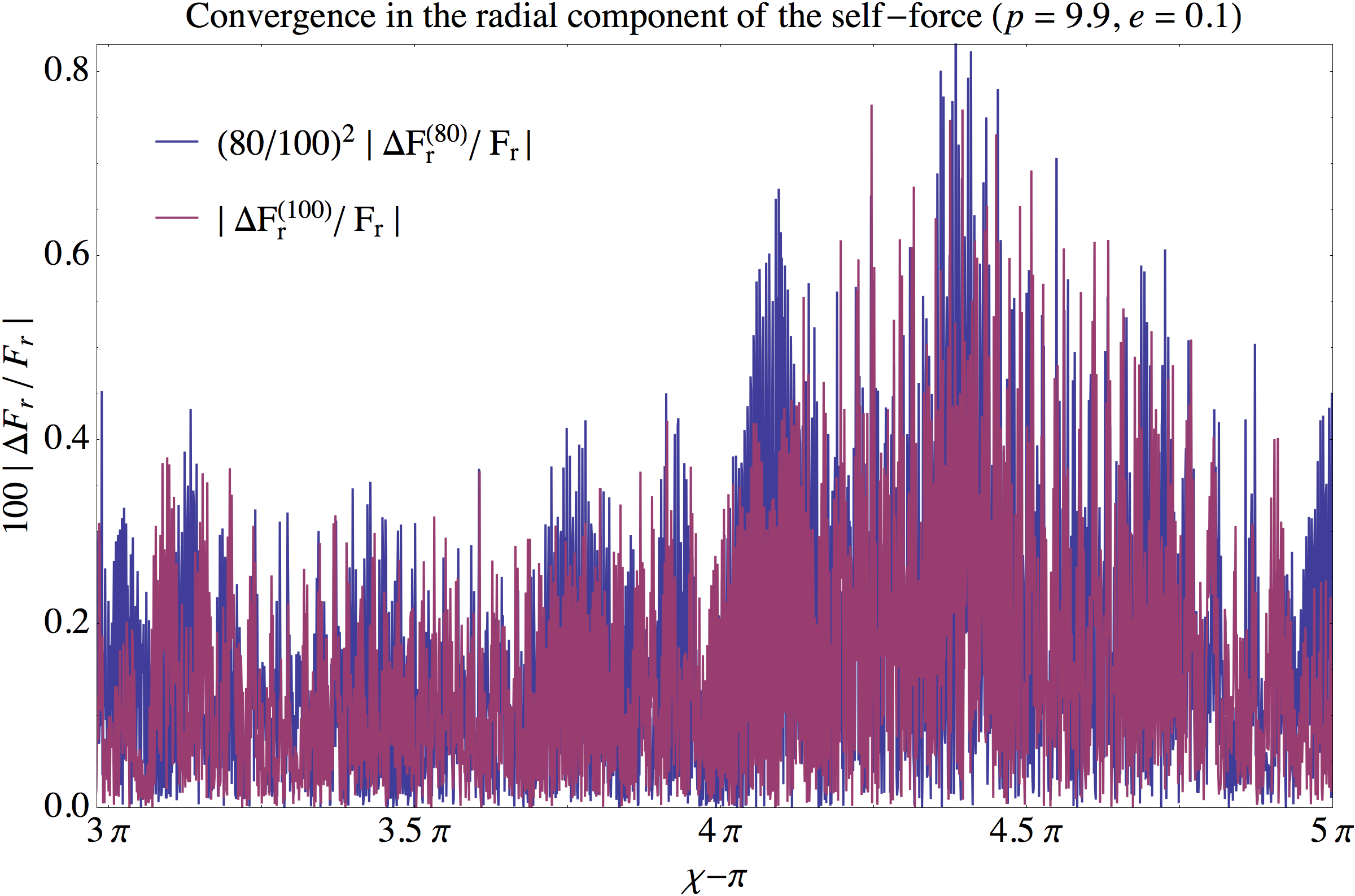}
\caption{Relative error in the $r$ component of the self-force for the $e=0.1$, $p=9.9$
case. When rescaled by the anticipated second-order convergence factor, the errors in the high
resolution simulation are comparable to those of the medium resolution.}
\label{fig:convergence-fr}
\end{center}
\end{figure}

Figures~\ref{fig:convergence-ft}, \ref{fig:convergence-fphi} and \ref{fig:convergence-fr} show
the convergence in $F_t$, $F_\phi$ and $F_r$ for the $e=0.1$, $p=9.9$ case by measuring errors
relative to reference values from the (1+1) time-domain code. At the medium and
high resolutions,
the error is dominated by the high-frequency errors coming from the low differentiability of
the solution near the world line and we see that the amplitude of the error converges away at
approximately second order, as expected.

In contrast, we found that our lowest resolution runs also contained
smooth finite differencing errors which scaled as the fifth power of the change in resolution.
This error arises simply because of insufficient resolution in the angular
directions (recall that our use of a window function in the polar direction introduces significant
angular structure). The increase in resolution to $80$ angular cells was sufficient to decrease
this error to below the level of the error arising from the nonsmoothness on the world line.

\section{Conclusion}

In this paper, we reported the successful extension of the effective source approach to the case of
eccentric orbits in the Schwarzschild geometry. This advance relied on many code adjustments, but principally on
the construction of a generic effective source as detailed in \cite{Wardell:2011gb}. Our code is
now capable of calculating the self-force to within of $1\%$ of the reference value for the
$t$- and $r$-components, and to within $0.1\%$ for the $\phi$-component. We have also shown
that at sufficiently high resolution our code is second-order convergent in the calculation of 
the self-force. This new code has been the basis of
the first self-consistent simulation of a self-forced orbit for a scalar charge
\cite{Diener:2011cc}. Finally, we have presented our self-force results in the form of ``loops",
which give the self-force components through one radial cycle of an eccentric orbit. This manner of
presenting eccentric-orbit self-force data makes some features apparent that are obscured when the
data is presented as standard time series.

In principle, the effective source method can also be adapted to handle a generic orbit in the Kerr
spacetime. The only essential challenge is the considerable additional complexity introduced in the
calculation of the effective source. We see this as the natural next step in this developing
research programme, for which results should be forthcoming.

\begin{acknowledgments}
The authors thank Niels Warburton, Norichika Sago, Eric Poisson, Steven Detweiler, and Frank
L\"offler for helpful comments and many fruitful discussions that helped shape this work. 
I. V. acknowledges partial financial support from the European Research Council under the European
UnionÕs Seventh Framework Programme (FP7/2007-2013)/ERC Grant No. 306425 ``Challenging General
Relativity'' and from the Marie Curie Career Integration Grant LIMITSOFGR-2011-TPS, and would like
to thank the hospitality of Jose Perico Esguerra and the National Institute of Physics, University
of the Philippines-Diliman, where parts of this manuscript were written.
B.W. gratefully acknowledges support from Science Foundation Ireland under Grant No.~10/RFP/PHY2847.
Portions
of this research were conducted with high performance computational resources provided by the
Louisiana Optical Network Initiative (http://www.loni.org/) and also used the Extreme Science and
Engineering Discovery Environment, which is supported by National Science Foundation Grant No.
OCI-1053575 (allocation TG-MCA02N014). The authors additionally wish to acknowledge the SFI/HEA
Irish Centre for High-End Computing (ICHEC) for the provision of computational facilities and
support (project ndast005b). Some computations were also performed on the Datura cluster at the
Albert Einstein Institute.
\end{acknowledgments}

\appendix*

\section{Flux formulas}
\label{app:flux}

In \cite{Vega:2009qb}, the expressions for the energy flux through the event horizon and a large spatial 2-sphere were derived. This appendix similarly derives the corresponding expressions for the angular momentum flux at the horizon ($\mathcal{H}$) in Kerr-Schild coordinates and at future null infinity ($\mathcal{J}^+$) in Cartesian hyperboloidal coordinates. 

Kerr-Schild and Schwarzschild coordinates are related according to
\begin{equation}
t = t_{KS} - 2M \ln\left(\frac{r}{2M}-1\right)
\end{equation} 
where $t$ is Schwarzschild time, $t_{KS}$ is Kerr-Schild time, and $r = (x^2+y^2+z^2)^{1/2}$ in Kerr-Schild coordinates $\{x,y,z\}$.

To implement hyperboloidal slicing (in the exterior region where the effective source vanishes, including $\scriP$), we use the additional transformation $\{t_{KS},r\}\rightarrow \{\tau,\rho\}$:
\begin{align}
\tau = t_{KS} - h(r)
\end{align}
\begin{align}
\frac{\rho}{\Omega(\rho)} = r
\end{align} 
where
the choices for $\Omega(\rho)$ and $h(r)$ in a neighborhood of $\scriP$ are the same as in \cite{Zenginoglu:2009hd,Vega:2011wf} (following the notation of
\cite{Zenginoglu:2009hd}):
\begin{equation}
\Omega(\rho) = 1- \frac{\rho}{\rho_{\scriP}}
\label{eqn:hyp1}
\end{equation}
\begin{equation}
\frac{dh}{dr} = 1 + \frac{4M\Omega}{\rho} + \frac{(8M^2-\rho^2_{\scriP})\Omega^2}{\rho^2},
\label{eqn:hyp2}
\end{equation}
so that $\scriP$ is located at $\rho = \rho_{\scriP}$. In this coordinate system, the metric is singular at $\scriP$, so we finally apply a conformal transformation, $\hat{g}_{\alpha\beta} = \Omega^2 g_{\alpha\beta}$. At $\scriP$, the conformal metric $\hat{g}_{\alpha\beta}$ is regular.

The angular momentum fluxes through $\mathcal{H}$ and $\mathcal{J}^+$
are respectively given by
\begin{align}
\left.\frac{dL}{dt}\right|_{\mathcal{H}} &= \oint_{H} \phi^\alpha T_{\alpha\beta} (-l^\beta) r^2\; d\Omega \label{eq:horizon}, \\
\left.\frac{dL}{d\tau}\right|_{\mathcal{J}^+} &= \oint_{\mathcal{J}^+} \phi^\alpha \hat{T}_{\alpha\beta} n^\beta \rho^2\; d\Omega. \label{eq:infty}
\end{align}
where 
\begin{equation}
T_{\alpha\beta} = \frac{1}{4\pi}\left(\nabla_\alpha\Phi\nabla_\beta\Phi - \frac{1}{2}g_{\alpha\beta}\nabla^\gamma\Phi\nabla_\gamma\Phi\right),
\label{eq:stressenergy}
\end{equation}
$\hat{T}_{\alpha\beta}$ is the stress-energy in the conformally-related space, $\phi^\alpha$ is the rotational Killing vector, while $l^\beta$ and $n^\beta$ are the null generators of $\mathcal{H}$ and $\mathcal{J}^+$, respectively.

Our goal is to write these flux formulas explicitly in terms of the quantities we compute in our code: the scalar field, $\Phi$, and its derivatives in Kerr-Schild and hyperboloidal coordinates. 

We shall deal with the angular momentum flux through $\mathcal{H}$ first. In  Kerr-Schild coordinates, the Schwarzschild metric and its inverse are simply
\begin{align}
g_{\alpha\beta} &= \eta_{\alpha\beta} + \frac{2M}{r}k_\alpha k_\beta, \label{eqn:KSmetric} \\
g^{\alpha\beta} &= \eta^{\alpha\beta} - \frac{2M}{r}k^\alpha k^\beta, \label{eqn:KSmetricinv}
\end{align}
\begin{equation}
k_\alpha = \left(1,\hat{n}_i\right), \,\,\,\,\,\,\, k^a = \left(1,-\hat{n}^i\right)  \label{eqn:nulls},
\end{equation}
where again $r=(x^2+y^2+z^2)^{1/2}$, $\hat{n}^i=x^i/r$, and $\eta_{\alpha\beta}=\mbox{diag}(-1,1,1,1)$.

The event horizon is essentially a surface of constant retarded time
$u = t-r-2M\ln{(r/2M-1)}$. In Kerr-Schild
coordinates these surfaces of constant $u$ are
\begin{equation}
t_{KS}=r+4M\ln{(r/2M-1)} + C,
\end{equation}
where $C$ is just a constant. 
In Kerr-Schild coordinates, the null generator of $\mathcal{H}$ is then just
\begin{equation}
l^\alpha_{\rm KS} = \delta^\alpha_{t_{KS}}.
\end{equation}
and the rotational Killing vector is 
\begin{equation}
\phi^\alpha_{\rm KS} = (0,-y,x,0).
\end{equation}
Putting everything together, we get
\begin{align}
T_{\alpha\beta}\phi^\alpha l^\beta = \frac{x\partial_y\Phi-y\partial_x\Phi}{4\pi} \frac{\partial\Phi}{\partial t_{KS}}. 
\label{eqn:Tphiel}
\end{align}
The angular momentum flux through the event horizon is then simply just
\begin{equation}
    \left.\frac{dL}{dt}\right|_{\mathcal{H}} = -\frac{M^2}{\pi}\oint_{r=2M} \frac{\partial\Phi}{\partial t_{KS}}\left(x\partial_y\Phi-y\partial_x\Phi\right) d\Omega.
\end{equation}\vspace{0.5em}

Now we turn to the flux through $\scriP$. The conformal metric close to $\scriP$ can be shown to be 
\begin{align}
d\hat{s}^2 = \hat{g}_{\alpha\beta}^{\rm hyp}dx^\alpha dx^\beta &:= (\Omega^2 g_{\alpha\beta}^{\rm hyp})dx^\alpha dx^\beta \nonumber \\ &\approx -2 d\tau d\rho + \rho^2_{\scriP} d\Omega^2
\end{align}
where we have used $\Omega(\rho) = 1-\rho/\rho_{\scriP}$ and $(dh/dr)|_{\scriP} = 1$, which follow from Eqs. (\ref{eqn:hyp1}) and (\ref{eqn:hyp2}).

The null generator of $\mathcal{J}^+$ is then
\begin{equation}
n^\alpha_{\rm hyp} = \hat{g}^{\alpha\beta}_{\rm hyp}\partial_\beta\rho= -\delta^\alpha_{\tau}.
\end{equation}

We can also switch to Cartesian hyperboloidal coordinates, $\{\hat{x},\hat{y},\hat{z} \}$, defined by
\begin{align}
\hat{x} &= \frac{\rho}{\Omega(\rho)}\sin\theta\cos\phi \\
\hat{y} &= \frac{\rho}{\Omega(\rho)}\sin\theta\sin\phi \\
\hat{z} &= \frac{\rho}{\Omega(\rho)}\cos\theta,
\end{align}
so that the rotational Killing vector becomes
\begin{equation}
\phi^\alpha_{\rm hyp} = (0,-\hat{y},\hat{x},0).
\end{equation}

We then find that
\begin{align}
\hat{T}_{\alpha\beta}\phi^\alpha n^\beta = \left(\frac{\hat{y}\partial_{\hat{x}}\hat{\Phi}-\hat{x}\partial_{\hat{y}}\hat{\Phi}}{4\pi}\right)\frac{\partial\hat{\Phi}}{\partial\tau},
\end{align}
which looks very similar to Eq. (\ref{eqn:Tphiel}), except that all the quantities here pertain to the conformally-related space, and not the physical space.

Finally we get 
\begin{equation}
\left.\frac{dL}{d\tau}\right|_{\mathcal{J}^+} = -\frac{\rho^2_{\scriP}}{4\pi}\oint_{\mathcal{J}^+} \frac{\partial\hat{\Phi}}{\partial\tau}\left(\hat{x}\partial_{\hat{y}}\hat{\Phi}-\hat{y}\partial_{\hat{x}}\hat{\Phi}\right)\; d\Omega.
\end{equation}

For completeness, we also include here an explicit expression for the energy flux through $\scriP$. In \cite{Vega:2009qb}, only the energy flux at spatial infinity was derived and was taken to be the limit of the flux through a spatial 2-sphere as the radius of the sphere approached infinity. With hyperboloidal slicing, the energy flux through $\scriP$ is just
\begin{align}
\left.\frac{dE}{d\tau}\right|_{\mathcal{J}^+} = \oint_{\mathcal{J}^+} t^\alpha \hat{T}_{\alpha\beta} n^\beta \rho^2\; d\Omega,
\end{align}
where $t^\alpha$ is just the timelike Killing vector of the Schwarzschild spacetime. In hyperboloidal coordinates, the timelike Killing vector also has components given by
\begin{equation}
t^\alpha_{\rm hyp} = \delta^\alpha_\tau.
\end{equation}
This then easily leads to the expression
\begin{equation}
\left.\frac{dE}{d\tau}\right|_{\mathcal{J}^+} = -\frac{\rho^2_{\scriP}}{4\pi}\oint_{\mathcal{J}^+} \left(\frac{\partial\hat{\Phi}}{\partial\tau}\right)^2\; d\Omega.
\end{equation}

\bibliography{myrefs}

\begin{thebibliography}{48}%
\makeatletter
\providecommand \@ifxundefined [1]{%
 \@ifx{#1\undefined}
}%
\providecommand \@ifnum [1]{%
 \ifnum #1\expandafter \@firstoftwo
 \else \expandafter \@secondoftwo
 \fi
}%
\providecommand \@ifx [1]{%
 \ifx #1\expandafter \@firstoftwo
 \else \expandafter \@secondoftwo
 \fi
}%
\providecommand \natexlab [1]{#1}%
\providecommand \enquote  [1]{``#1''}%
\providecommand \bibnamefont  [1]{#1}%
\providecommand \bibfnamefont [1]{#1}%
\providecommand \citenamefont [1]{#1}%
\providecommand \href@noop [0]{\@secondoftwo}%
\providecommand \href [0]{\begingroup \@sanitize@url \@href}%
\providecommand \@href[1]{\@@startlink{#1}\@@href}%
\providecommand \@@href[1]{\endgroup#1\@@endlink}%
\providecommand \@sanitize@url [0]{\catcode `\\12\catcode `\$12\catcode
  `\&12\catcode `\#12\catcode `\^12\catcode `\_12\catcode `\%12\relax}%
\providecommand \@@startlink[1]{}%
\providecommand \@@endlink[0]{}%
\providecommand \url  [0]{\begingroup\@sanitize@url \@url }%
\providecommand \@url [1]{\endgroup\@href {#1}{\urlprefix }}%
\providecommand \urlprefix  [0]{URL }%
\providecommand \Eprint [0]{\href }%
\providecommand \doibase [0]{http://dx.doi.org/}%
\providecommand \selectlanguage [0]{\@gobble}%
\providecommand \bibinfo  [0]{\@secondoftwo}%
\providecommand \bibfield  [0]{\@secondoftwo}%
\providecommand \translation [1]{[#1]}%
\providecommand \BibitemOpen [0]{}%
\providecommand \bibitemStop [0]{}%
\providecommand \bibitemNoStop [0]{.\EOS\space}%
\providecommand \EOS [0]{\spacefactor3000\relax}%
\providecommand \BibitemShut  [1]{\csname bibitem#1\endcsname}%
\let\auto@bib@innerbib\@empty
\bibitem [{\citenamefont {Sathyaprakash}\ and\ \citenamefont
  {Schutz}(2009)}]{Sathyaprakash:2009xs}%
  \BibitemOpen
  \bibfield  {author} {\bibinfo {author} {\bibfnamefont {B.}~\bibnamefont
  {Sathyaprakash}}\ and\ \bibinfo {author} {\bibfnamefont {B.}~\bibnamefont
  {Schutz}},\ }\href@noop {} {\bibfield  {journal} {\bibinfo  {journal} {Living
  Rev.Rel.}\ }\textbf {\bibinfo {volume} {12}},\ \bibinfo {pages} {2} (\bibinfo
  {year} {2009})},\ \Eprint {http://arxiv.org/abs/0903.0338} {arXiv:0903.0338
  [gr-qc]} \BibitemShut {NoStop}%
\bibitem [{\citenamefont {Amaro-Seoane}\ \emph {et~al.}(2010)\citenamefont
  {Amaro-Seoane}, \citenamefont {Schutz},\ and\ \citenamefont
  {Sopuerta}}]{AmaroSeoane:2010zy}%
  \BibitemOpen
  \bibfield  {author} {\bibinfo {author} {\bibfnamefont {P.}~\bibnamefont
  {Amaro-Seoane}}, \bibinfo {author} {\bibfnamefont {B.}~\bibnamefont
  {Schutz}}, \ and\ \bibinfo {author} {\bibfnamefont {C.~F.}\ \bibnamefont
  {Sopuerta}},\ }\href@noop {} {\  (\bibinfo {year} {2010})},\ \Eprint
  {http://arxiv.org/abs/1009.1402} {arXiv:1009.1402 [astro-ph.CO]} \BibitemShut
  {NoStop}%
\bibitem [{\citenamefont {Amaro-Seoane}\ \emph {et~al.}(2013)\citenamefont
  {Amaro-Seoane}, \citenamefont {Aoudia}, \citenamefont {Babak}, \citenamefont
  {Binetruy}, \citenamefont {Berti} \emph {et~al.}}]{AmaroSeoane:2012km}%
  \BibitemOpen
  \bibfield  {author} {\bibinfo {author} {\bibfnamefont {P.}~\bibnamefont
  {Amaro-Seoane}}, \bibinfo {author} {\bibfnamefont {S.}~\bibnamefont
  {Aoudia}}, \bibinfo {author} {\bibfnamefont {S.}~\bibnamefont {Babak}},
  \bibinfo {author} {\bibfnamefont {P.}~\bibnamefont {Binetruy}}, \bibinfo
  {author} {\bibfnamefont {E.}~\bibnamefont {Berti}},  \emph {et~al.},\
  }\href@noop {} {\bibfield  {journal} {\bibinfo  {journal} {GW Notes}\
  }\textbf {\bibinfo {volume} {6}} (\bibinfo {year} {2013})},\ \Eprint
  {http://arxiv.org/abs/1201.3621} {arXiv:1201.3621 [astro-ph.CO]} \BibitemShut
  {NoStop}%
\bibitem [{\citenamefont {Davis}\ \emph {et~al.}(1971)\citenamefont {Davis},
  \citenamefont {Ruffini}, \citenamefont {Press},\ and\ \citenamefont
  {Price}}]{davis-etal:71}%
  \BibitemOpen
  \bibfield  {author} {\bibinfo {author} {\bibfnamefont {M.}~\bibnamefont
  {Davis}}, \bibinfo {author} {\bibfnamefont {R.}~\bibnamefont {Ruffini}},
  \bibinfo {author} {\bibfnamefont {W.}~\bibnamefont {Press}}, \ and\ \bibinfo
  {author} {\bibfnamefont {R.}~\bibnamefont {Price}},\ }\href@noop {}
  {\bibfield  {journal} {\bibinfo  {journal} {Phys. Rev. Lett.}\ }\textbf
  {\bibinfo {volume} {27}},\ \bibinfo {pages} {1466} (\bibinfo {year}
  {1971})}\BibitemShut {NoStop}%
\bibitem [{\citenamefont {Detweiler}(1978)}]{detweiler:78}%
  \BibitemOpen
  \bibfield  {author} {\bibinfo {author} {\bibfnamefont {S.}~\bibnamefont
  {Detweiler}},\ }\href@noop {} {\bibfield  {journal} {\bibinfo  {journal}
  {Astrophys. J.}\ }\textbf {\bibinfo {volume} {225}},\ \bibinfo {pages} {687}
  (\bibinfo {year} {1978})}\BibitemShut {NoStop}%
\bibitem [{\citenamefont {Barack}\ and\ \citenamefont
  {Golbourn}(2007)}]{Barack:2007jh}%
  \BibitemOpen
  \bibfield  {author} {\bibinfo {author} {\bibfnamefont {L.}~\bibnamefont
  {Barack}}\ and\ \bibinfo {author} {\bibfnamefont {D.~A.}\ \bibnamefont
  {Golbourn}},\ }\href {\doibase 10.1103/PhysRevD.76.044020} {\bibfield
  {journal} {\bibinfo  {journal} {Phys.Rev.}\ }\textbf {\bibinfo {volume}
  {D76}},\ \bibinfo {pages} {044020} (\bibinfo {year} {2007})},\ \Eprint
  {http://arxiv.org/abs/0705.3620} {arXiv:0705.3620 [gr-qc]} \BibitemShut
  {NoStop}%
\bibitem [{\citenamefont {Vega}\ and\ \citenamefont
  {Detweiler}(2008)}]{Vega:2007mc}%
  \BibitemOpen
  \bibfield  {author} {\bibinfo {author} {\bibfnamefont {I.}~\bibnamefont
  {Vega}}\ and\ \bibinfo {author} {\bibfnamefont {S.~L.}\ \bibnamefont
  {Detweiler}},\ }\href {\doibase 10.1103/PhysRevD.77.084008} {\bibfield
  {journal} {\bibinfo  {journal} {Phys.Rev.}\ }\textbf {\bibinfo {volume}
  {D77}},\ \bibinfo {pages} {084008} (\bibinfo {year} {2008})},\ \Eprint
  {http://arxiv.org/abs/0712.4405} {arXiv:0712.4405 [gr-qc]} \BibitemShut
  {NoStop}%
\bibitem [{\citenamefont {Diener}\ \emph {et~al.}(2012)\citenamefont {Diener},
  \citenamefont {Vega}, \citenamefont {Wardell},\ and\ \citenamefont
  {Detweiler}}]{Diener:2011cc}%
  \BibitemOpen
  \bibfield  {author} {\bibinfo {author} {\bibfnamefont {P.}~\bibnamefont
  {Diener}}, \bibinfo {author} {\bibfnamefont {I.}~\bibnamefont {Vega}},
  \bibinfo {author} {\bibfnamefont {B.}~\bibnamefont {Wardell}}, \ and\
  \bibinfo {author} {\bibfnamefont {S.}~\bibnamefont {Detweiler}},\ }\href
  {\doibase 10.1103/PhysRevLett.108.191102} {\bibfield  {journal} {\bibinfo
  {journal} {Phys.Rev.Lett.}\ }\textbf {\bibinfo {volume} {108}},\ \bibinfo
  {pages} {191102} (\bibinfo {year} {2012})},\ \Eprint
  {http://arxiv.org/abs/1112.4821} {arXiv:1112.4821 [gr-qc]} \BibitemShut
  {NoStop}%
\bibitem [{\citenamefont {Vega}\ \emph {et~al.}(2009)\citenamefont {Vega},
  \citenamefont {Diener}, \citenamefont {Tichy},\ and\ \citenamefont
  {Detweiler}}]{Vega:2009qb}%
  \BibitemOpen
  \bibfield  {author} {\bibinfo {author} {\bibfnamefont {I.}~\bibnamefont
  {Vega}}, \bibinfo {author} {\bibfnamefont {P.}~\bibnamefont {Diener}},
  \bibinfo {author} {\bibfnamefont {W.}~\bibnamefont {Tichy}}, \ and\ \bibinfo
  {author} {\bibfnamefont {S.~L.}\ \bibnamefont {Detweiler}},\ }\href {\doibase
  10.1103/PhysRevD.80.084021} {\bibfield  {journal} {\bibinfo  {journal}
  {Phys.Rev.}\ }\textbf {\bibinfo {volume} {D80}},\ \bibinfo {pages} {084021}
  (\bibinfo {year} {2009})},\ \Eprint {http://arxiv.org/abs/0908.2138}
  {arXiv:0908.2138 [gr-qc]} \BibitemShut {NoStop}%
\bibitem [{\citenamefont {Wardell}\ \emph {et~al.}(2012)\citenamefont
  {Wardell}, \citenamefont {Vega}, \citenamefont {Thornburg},\ and\
  \citenamefont {Diener}}]{Wardell:2011gb}%
  \BibitemOpen
  \bibfield  {author} {\bibinfo {author} {\bibfnamefont {B.}~\bibnamefont
  {Wardell}}, \bibinfo {author} {\bibfnamefont {I.}~\bibnamefont {Vega}},
  \bibinfo {author} {\bibfnamefont {J.}~\bibnamefont {Thornburg}}, \ and\
  \bibinfo {author} {\bibfnamefont {P.}~\bibnamefont {Diener}},\ }\href
  {\doibase 10.1103/PhysRevD.85.104044} {\bibfield  {journal} {\bibinfo
  {journal} {Phys.Rev.}\ }\textbf {\bibinfo {volume} {D85}},\ \bibinfo {pages}
  {104044} (\bibinfo {year} {2012})},\ \Eprint {http://arxiv.org/abs/1112.6355}
  {arXiv:1112.6355 [gr-qc]} \BibitemShut {NoStop}%
\bibitem [{\citenamefont {Haas}(2007)}]{Haas:2007kz}%
  \BibitemOpen
  \bibfield  {author} {\bibinfo {author} {\bibfnamefont {R.}~\bibnamefont
  {Haas}},\ }\href {\doibase 10.1103/PhysRevD.75.124011} {\bibfield  {journal}
  {\bibinfo  {journal} {Phys.Rev.}\ }\textbf {\bibinfo {volume} {D75}},\
  \bibinfo {pages} {124011} (\bibinfo {year} {2007})},\ \Eprint
  {http://arxiv.org/abs/0704.0797} {arXiv:0704.0797 [gr-qc]} \BibitemShut
  {NoStop}%
\bibitem [{\citenamefont {Misner}\ \emph {et~al.}(1974)\citenamefont {Misner},
  \citenamefont {Thorne},\ and\ \citenamefont {Wheeler}}]{Misner:1974qy}%
  \BibitemOpen
  \bibfield  {author} {\bibinfo {author} {\bibfnamefont {C.~W.}\ \bibnamefont
  {Misner}}, \bibinfo {author} {\bibfnamefont {K.}~\bibnamefont {Thorne}}, \
  and\ \bibinfo {author} {\bibfnamefont {J.}~\bibnamefont {Wheeler}},\
  }\href@noop {} {\emph {\bibinfo {title} {{Gravitation}}}}\ (\bibinfo
  {publisher} {Freeman},\ \bibinfo {address} {San Francisco},\ \bibinfo {year}
  {1974})\BibitemShut {NoStop}%
\bibitem [{\citenamefont {Cutler}\ \emph {et~al.}(1994)\citenamefont {Cutler},
  \citenamefont {Kennefick},\ and\ \citenamefont {Poisson}}]{Cutler:1994pb}%
  \BibitemOpen
  \bibfield  {author} {\bibinfo {author} {\bibfnamefont {C.}~\bibnamefont
  {Cutler}}, \bibinfo {author} {\bibfnamefont {D.}~\bibnamefont {Kennefick}}, \
  and\ \bibinfo {author} {\bibfnamefont {E.}~\bibnamefont {Poisson}},\ }\href
  {\doibase 10.1103/PhysRevD.50.3816} {\bibfield  {journal} {\bibinfo
  {journal} {Phys.Rev.}\ }\textbf {\bibinfo {volume} {D50}},\ \bibinfo {pages}
  {3816} (\bibinfo {year} {1994})}\BibitemShut {NoStop}%
\bibitem [{\citenamefont {Pound}\ and\ \citenamefont
  {Poisson}(2008)}]{Pound:2007th}%
  \BibitemOpen
  \bibfield  {author} {\bibinfo {author} {\bibfnamefont {A.}~\bibnamefont
  {Pound}}\ and\ \bibinfo {author} {\bibfnamefont {E.}~\bibnamefont
  {Poisson}},\ }\href {\doibase 10.1103/PhysRevD.77.044013} {\bibfield
  {journal} {\bibinfo  {journal} {Phys.Rev.}\ }\textbf {\bibinfo {volume}
  {D77}},\ \bibinfo {pages} {044013} (\bibinfo {year} {2008})},\ \Eprint
  {http://arxiv.org/abs/0708.3033} {arXiv:0708.3033 [gr-qc]} \BibitemShut
  {NoStop}%
\bibitem [{\citenamefont {Barack}\ and\ \citenamefont
  {Sago}(2010)}]{Barack:2010tm}%
  \BibitemOpen
  \bibfield  {author} {\bibinfo {author} {\bibfnamefont {L.}~\bibnamefont
  {Barack}}\ and\ \bibinfo {author} {\bibfnamefont {N.}~\bibnamefont {Sago}},\
  }\href {\doibase 10.1103/PhysRevD.81.084021} {\bibfield  {journal} {\bibinfo
  {journal} {Phys.Rev.}\ }\textbf {\bibinfo {volume} {D81}},\ \bibinfo {pages}
  {084021} (\bibinfo {year} {2010})},\ \Eprint {http://arxiv.org/abs/1002.2386}
  {arXiv:1002.2386 [gr-qc]} \BibitemShut {NoStop}%
\bibitem [{\citenamefont {Quinn}(2000)}]{Quinn:2000wa}%
  \BibitemOpen
  \bibfield  {author} {\bibinfo {author} {\bibfnamefont {T.~C.}\ \bibnamefont
  {Quinn}},\ }\href {\doibase 10.1103/PhysRevD.62.064029} {\bibfield  {journal}
  {\bibinfo  {journal} {Phys.Rev.}\ }\textbf {\bibinfo {volume} {D62}},\
  \bibinfo {pages} {064029} (\bibinfo {year} {2000})},\ \Eprint
  {http://arxiv.org/abs/gr-qc/0005030} {arXiv:gr-qc/0005030 [gr-qc]}
  \BibitemShut {NoStop}%
\bibitem [{\citenamefont {Haas}(2011)}]{Haas:2011np}%
  \BibitemOpen
  \bibfield  {author} {\bibinfo {author} {\bibfnamefont {R.}~\bibnamefont
  {Haas}},\ }\href@noop {} {\  (\bibinfo {year} {2011})},\ \Eprint
  {http://arxiv.org/abs/1112.3707} {arXiv:1112.3707 [gr-qc]} \BibitemShut
  {NoStop}%
\bibitem [{\citenamefont {Canizares}\ and\ \citenamefont
  {Sopuerta}(2009)}]{Canizares:2009ay}%
  \BibitemOpen
  \bibfield  {author} {\bibinfo {author} {\bibfnamefont {P.}~\bibnamefont
  {Canizares}}\ and\ \bibinfo {author} {\bibfnamefont {C.~F.}\ \bibnamefont
  {Sopuerta}},\ }\href {\doibase 10.1103/PhysRevD.79.084020} {\bibfield
  {journal} {\bibinfo  {journal} {Phys.Rev.}\ }\textbf {\bibinfo {volume}
  {D79}},\ \bibinfo {pages} {084020} (\bibinfo {year} {2009})},\ \Eprint
  {http://arxiv.org/abs/0903.0505} {arXiv:0903.0505 [gr-qc]} \BibitemShut
  {NoStop}%
\bibitem [{\citenamefont {Canizares}\ \emph {et~al.}(2010)\citenamefont
  {Canizares}, \citenamefont {Sopuerta},\ and\ \citenamefont
  {Jaramillo}}]{Canizares:2010yx}%
  \BibitemOpen
  \bibfield  {author} {\bibinfo {author} {\bibfnamefont {P.}~\bibnamefont
  {Canizares}}, \bibinfo {author} {\bibfnamefont {C.~F.}\ \bibnamefont
  {Sopuerta}}, \ and\ \bibinfo {author} {\bibfnamefont {J.~L.}\ \bibnamefont
  {Jaramillo}},\ }\href {\doibase 10.1103/PhysRevD.82.044023} {\bibfield
  {journal} {\bibinfo  {journal} {Phys.Rev.}\ }\textbf {\bibinfo {volume}
  {D82}},\ \bibinfo {pages} {044023} (\bibinfo {year} {2010})},\ \Eprint
  {http://arxiv.org/abs/1006.3201} {arXiv:1006.3201 [gr-qc]} \BibitemShut
  {NoStop}%
\bibitem [{\citenamefont {Canizares}\ and\ \citenamefont
  {Sopuerta}(2011)}]{Canizares:2011kw}%
  \BibitemOpen
  \bibfield  {author} {\bibinfo {author} {\bibfnamefont {P.}~\bibnamefont
  {Canizares}}\ and\ \bibinfo {author} {\bibfnamefont {C.~F.}\ \bibnamefont
  {Sopuerta}},\ }\href {\doibase 10.1088/0264-9381/28/13/134011} {\bibfield
  {journal} {\bibinfo  {journal} {Class.Quant.Grav.}\ }\textbf {\bibinfo
  {volume} {28}},\ \bibinfo {pages} {134011} (\bibinfo {year} {2011})},\
  \Eprint {http://arxiv.org/abs/1101.2526} {arXiv:1101.2526 [gr-qc]}
  \BibitemShut {NoStop}%
\bibitem [{\citenamefont {Warburton}\ and\ \citenamefont
  {Barack}(2010)}]{Warburton:2010eq}%
  \BibitemOpen
  \bibfield  {author} {\bibinfo {author} {\bibfnamefont {N.}~\bibnamefont
  {Warburton}}\ and\ \bibinfo {author} {\bibfnamefont {L.}~\bibnamefont
  {Barack}},\ }\href {\doibase 10.1103/PhysRevD.81.084039} {\bibfield
  {journal} {\bibinfo  {journal} {Phys.Rev.}\ }\textbf {\bibinfo {volume}
  {D81}},\ \bibinfo {pages} {084039} (\bibinfo {year} {2010})},\ \Eprint
  {http://arxiv.org/abs/1003.1860} {arXiv:1003.1860 [gr-qc]} \BibitemShut
  {NoStop}%
\bibitem [{\citenamefont {Warburton}\ and\ \citenamefont
  {Barack}(2011)}]{Warburton:2011hp}%
  \BibitemOpen
  \bibfield  {author} {\bibinfo {author} {\bibfnamefont {N.}~\bibnamefont
  {Warburton}}\ and\ \bibinfo {author} {\bibfnamefont {L.}~\bibnamefont
  {Barack}},\ }\href {\doibase 10.1103/PhysRevD.83.124038} {\bibfield
  {journal} {\bibinfo  {journal} {Phys.Rev.}\ }\textbf {\bibinfo {volume}
  {D83}},\ \bibinfo {pages} {124038} (\bibinfo {year} {2011})},\ \Eprint
  {http://arxiv.org/abs/1103.0287} {arXiv:1103.0287 [gr-qc]} \BibitemShut
  {NoStop}%
\bibitem [{\citenamefont {Zimmerman}\ \emph {et~al.}(2013)\citenamefont
  {Zimmerman}, \citenamefont {Vega}, \citenamefont {Poisson},\ and\
  \citenamefont {Haas}}]{Zimmerman:2012zu}%
  \BibitemOpen
  \bibfield  {author} {\bibinfo {author} {\bibfnamefont {P.}~\bibnamefont
  {Zimmerman}}, \bibinfo {author} {\bibfnamefont {I.}~\bibnamefont {Vega}},
  \bibinfo {author} {\bibfnamefont {E.}~\bibnamefont {Poisson}}, \ and\
  \bibinfo {author} {\bibfnamefont {R.}~\bibnamefont {Haas}},\ }\href {\doibase
  10.1103/PhysRevD.87.041501} {\bibfield  {journal} {\bibinfo  {journal}
  {Phys.Rev.}\ }\textbf {\bibinfo {volume} {D87}},\ \bibinfo {pages}
  {041501(R)} (\bibinfo {year} {2013})},\ \Eprint
  {http://arxiv.org/abs/1211.3889} {arXiv:1211.3889 [gr-qc]} \BibitemShut
  {NoStop}%
\bibitem [{\citenamefont {Warburton}\ \emph {et~al.}(2012)\citenamefont
  {Warburton}, \citenamefont {Akcay}, \citenamefont {Barack}, \citenamefont
  {Gair},\ and\ \citenamefont {Sago}}]{Warburton:2011fk}%
  \BibitemOpen
  \bibfield  {author} {\bibinfo {author} {\bibfnamefont {N.}~\bibnamefont
  {Warburton}}, \bibinfo {author} {\bibfnamefont {S.}~\bibnamefont {Akcay}},
  \bibinfo {author} {\bibfnamefont {L.}~\bibnamefont {Barack}}, \bibinfo
  {author} {\bibfnamefont {J.~R.}\ \bibnamefont {Gair}}, \ and\ \bibinfo
  {author} {\bibfnamefont {N.}~\bibnamefont {Sago}},\ }\href {\doibase
  10.1103/PhysRevD.85.061501} {\bibfield  {journal} {\bibinfo  {journal}
  {Phys.Rev.}\ }\textbf {\bibinfo {volume} {D85}},\ \bibinfo {pages} {061501}
  (\bibinfo {year} {2012})},\ \Eprint {http://arxiv.org/abs/1111.6908}
  {arXiv:1111.6908 [gr-qc]} \BibitemShut {NoStop}%
\bibitem [{\citenamefont {Hughes}\ \emph {et~al.}(2005)\citenamefont {Hughes},
  \citenamefont {Drasco}, \citenamefont {Flanagan},\ and\ \citenamefont
  {Franklin}}]{Hughes:2005qb}%
  \BibitemOpen
  \bibfield  {author} {\bibinfo {author} {\bibfnamefont {S.~A.}\ \bibnamefont
  {Hughes}}, \bibinfo {author} {\bibfnamefont {S.}~\bibnamefont {Drasco}},
  \bibinfo {author} {\bibfnamefont {E.~E.}\ \bibnamefont {Flanagan}}, \ and\
  \bibinfo {author} {\bibfnamefont {J.}~\bibnamefont {Franklin}},\ }\href
  {\doibase 10.1103/PhysRevLett.94.221101} {\bibfield  {journal} {\bibinfo
  {journal} {Phys.Rev.Lett.}\ }\textbf {\bibinfo {volume} {94}},\ \bibinfo
  {pages} {221101} (\bibinfo {year} {2005})},\ \Eprint
  {http://arxiv.org/abs/gr-qc/0504015} {arXiv:gr-qc/0504015 [gr-qc]}
  \BibitemShut {NoStop}%
\bibitem [{\citenamefont {Drasco}\ \emph {et~al.}(2005)\citenamefont {Drasco},
  \citenamefont {Flanagan},\ and\ \citenamefont {Hughes}}]{Drasco:2005is}%
  \BibitemOpen
  \bibfield  {author} {\bibinfo {author} {\bibfnamefont {S.}~\bibnamefont
  {Drasco}}, \bibinfo {author} {\bibfnamefont {E.~E.}\ \bibnamefont
  {Flanagan}}, \ and\ \bibinfo {author} {\bibfnamefont {S.~A.}\ \bibnamefont
  {Hughes}},\ }\href {\doibase 10.1088/0264-9381/22/15/011} {\bibfield
  {journal} {\bibinfo  {journal} {Class.Quant.Grav.}\ }\textbf {\bibinfo
  {volume} {22}},\ \bibinfo {pages} {S801} (\bibinfo {year} {2005})},\ \Eprint
  {http://arxiv.org/abs/gr-qc/0505075} {arXiv:gr-qc/0505075 [gr-qc]}
  \BibitemShut {NoStop}%
\bibitem [{\citenamefont {Pound}\ \emph {et~al.}(2005)\citenamefont {Pound},
  \citenamefont {Poisson},\ and\ \citenamefont {Nickel}}]{Pound:2005fs}%
  \BibitemOpen
  \bibfield  {author} {\bibinfo {author} {\bibfnamefont {A.}~\bibnamefont
  {Pound}}, \bibinfo {author} {\bibfnamefont {E.}~\bibnamefont {Poisson}}, \
  and\ \bibinfo {author} {\bibfnamefont {B.~G.}\ \bibnamefont {Nickel}},\
  }\href {\doibase 10.1103/PhysRevD.72.124001} {\bibfield  {journal} {\bibinfo
  {journal} {Phys.Rev.}\ }\textbf {\bibinfo {volume} {D72}},\ \bibinfo {pages}
  {124001} (\bibinfo {year} {2005})},\ \Eprint
  {http://arxiv.org/abs/gr-qc/0509122} {arXiv:gr-qc/0509122 [gr-qc]}
  \BibitemShut {NoStop}%
\bibitem [{\citenamefont {Hinderer}\ and\ \citenamefont
  {Flanagan}(2008)}]{Hinderer:2008dm}%
  \BibitemOpen
  \bibfield  {author} {\bibinfo {author} {\bibfnamefont {T.}~\bibnamefont
  {Hinderer}}\ and\ \bibinfo {author} {\bibfnamefont {E.~E.}\ \bibnamefont
  {Flanagan}},\ }\href {\doibase 10.1103/PhysRevD.78.064028} {\bibfield
  {journal} {\bibinfo  {journal} {Phys.Rev.}\ }\textbf {\bibinfo {volume}
  {D78}},\ \bibinfo {pages} {064028} (\bibinfo {year} {2008})},\ \Eprint
  {http://arxiv.org/abs/0805.3337} {arXiv:0805.3337 [gr-qc]} \BibitemShut
  {NoStop}%
\bibitem [{\citenamefont {Mino}(2003)}]{Mino:2003yg}%
  \BibitemOpen
  \bibfield  {author} {\bibinfo {author} {\bibfnamefont {Y.}~\bibnamefont
  {Mino}},\ }\href {\doibase 10.1103/PhysRevD.67.084027} {\bibfield  {journal}
  {\bibinfo  {journal} {Phys.Rev.}\ }\textbf {\bibinfo {volume} {D67}},\
  \bibinfo {pages} {084027} (\bibinfo {year} {2003})},\ \Eprint
  {http://arxiv.org/abs/gr-qc/0302075} {arXiv:gr-qc/0302075 [gr-qc]}
  \BibitemShut {NoStop}%
\bibitem [{\citenamefont {Barack}(2009)}]{Barack:2009ux}%
  \BibitemOpen
  \bibfield  {author} {\bibinfo {author} {\bibfnamefont {L.}~\bibnamefont
  {Barack}},\ }\href {\doibase 10.1088/0264-9381/26/21/213001} {\bibfield
  {journal} {\bibinfo  {journal} {Class.Quant.Grav.}\ }\textbf {\bibinfo
  {volume} {26}},\ \bibinfo {pages} {213001} (\bibinfo {year} {2009})},\
  \Eprint {http://arxiv.org/abs/0908.1664} {arXiv:0908.1664 [gr-qc]}
  \BibitemShut {NoStop}%
\bibitem [{\citenamefont {Pfenning}\ and\ \citenamefont
  {Poisson}(2002)}]{Pfenning:2000zf}%
  \BibitemOpen
  \bibfield  {author} {\bibinfo {author} {\bibfnamefont {M.~J.}\ \bibnamefont
  {Pfenning}}\ and\ \bibinfo {author} {\bibfnamefont {E.}~\bibnamefont
  {Poisson}},\ }\href {\doibase 10.1103/PhysRevD.65.084001} {\bibfield
  {journal} {\bibinfo  {journal} {Phys.Rev.}\ }\textbf {\bibinfo {volume}
  {D65}},\ \bibinfo {pages} {084001} (\bibinfo {year} {2002})},\ \Eprint
  {http://arxiv.org/abs/gr-qc/0012057} {arXiv:gr-qc/0012057 [gr-qc]}
  \BibitemShut {NoStop}%
\bibitem [{\citenamefont {Vega}\ \emph {et~al.}(2011)\citenamefont {Vega},
  \citenamefont {Wardell},\ and\ \citenamefont {Diener}}]{Vega:2011wf}%
  \BibitemOpen
  \bibfield  {author} {\bibinfo {author} {\bibfnamefont {I.}~\bibnamefont
  {Vega}}, \bibinfo {author} {\bibfnamefont {B.}~\bibnamefont {Wardell}}, \
  and\ \bibinfo {author} {\bibfnamefont {P.}~\bibnamefont {Diener}},\ }\href
  {\doibase 10.1088/0264-9381/28/13/134010} {\bibfield  {journal} {\bibinfo
  {journal} {Class.Quant.Grav.}\ }\textbf {\bibinfo {volume} {28}},\ \bibinfo
  {pages} {134010} (\bibinfo {year} {2011})},\ \Eprint
  {http://arxiv.org/abs/1101.2925} {arXiv:1101.2925 [gr-qc]} \BibitemShut
  {NoStop}%
\bibitem [{\citenamefont {Zenginoglu}(2008)}]{Zenginoglu:2007jw}%
  \BibitemOpen
  \bibfield  {author} {\bibinfo {author} {\bibfnamefont {A.}~\bibnamefont
  {Zenginoglu}},\ }\href {\doibase 10.1088/0264-9381/25/14/145002} {\bibfield
  {journal} {\bibinfo  {journal} {Class.Quant.Grav.}\ }\textbf {\bibinfo
  {volume} {25}},\ \bibinfo {pages} {145002} (\bibinfo {year} {2008})},\
  \Eprint {http://arxiv.org/abs/0712.4333} {arXiv:0712.4333 [gr-qc]}
  \BibitemShut {NoStop}%
\bibitem [{\citenamefont {Zenginoglu}\ and\ \citenamefont
  {Tiglio}(2009)}]{Zenginoglu:2009hd}%
  \BibitemOpen
  \bibfield  {author} {\bibinfo {author} {\bibfnamefont {A.}~\bibnamefont
  {Zenginoglu}}\ and\ \bibinfo {author} {\bibfnamefont {M.}~\bibnamefont
  {Tiglio}},\ }\href {\doibase 10.1103/PhysRevD.80.024044} {\bibfield
  {journal} {\bibinfo  {journal} {Phys.Rev.}\ }\textbf {\bibinfo {volume}
  {D80}},\ \bibinfo {pages} {024044} (\bibinfo {year} {2009})},\ \Eprint
  {http://arxiv.org/abs/0906.3342} {arXiv:0906.3342 [gr-qc]} \BibitemShut
  {NoStop}%
\bibitem [{\citenamefont {Detweiler}\ and\ \citenamefont
  {Whiting}(2003)}]{Detweiler:2002mi}%
  \BibitemOpen
  \bibfield  {author} {\bibinfo {author} {\bibfnamefont {S.~L.}\ \bibnamefont
  {Detweiler}}\ and\ \bibinfo {author} {\bibfnamefont {B.~F.}\ \bibnamefont
  {Whiting}},\ }\href {\doibase 10.1103/PhysRevD.67.024025} {\bibfield
  {journal} {\bibinfo  {journal} {Phys.Rev.}\ }\textbf {\bibinfo {volume}
  {D67}},\ \bibinfo {pages} {024025} (\bibinfo {year} {2003})},\ \Eprint
  {http://arxiv.org/abs/gr-qc/0202086} {arXiv:gr-qc/0202086 [gr-qc]}
  \BibitemShut {NoStop}%
\bibitem [{\citenamefont {Dolan}\ and\ \citenamefont
  {Barack}(2011)}]{Dolan:2010mt}%
  \BibitemOpen
  \bibfield  {author} {\bibinfo {author} {\bibfnamefont {S.~R.}\ \bibnamefont
  {Dolan}}\ and\ \bibinfo {author} {\bibfnamefont {L.}~\bibnamefont {Barack}},\
  }\href {\doibase 10.1103/PhysRevD.83.024019} {\bibfield  {journal} {\bibinfo
  {journal} {Phys.Rev.}\ }\textbf {\bibinfo {volume} {D83}},\ \bibinfo {pages}
  {024019} (\bibinfo {year} {2011})},\ \Eprint {http://arxiv.org/abs/1010.5255}
  {arXiv:1010.5255 [gr-qc]} \BibitemShut {NoStop}%
\bibitem [{\citenamefont {Dolan}\ \emph {et~al.}(2011)\citenamefont {Dolan},
  \citenamefont {Barack},\ and\ \citenamefont {Wardell}}]{Dolan:2011dx}%
  \BibitemOpen
  \bibfield  {author} {\bibinfo {author} {\bibfnamefont {S.~R.}\ \bibnamefont
  {Dolan}}, \bibinfo {author} {\bibfnamefont {L.}~\bibnamefont {Barack}}, \
  and\ \bibinfo {author} {\bibfnamefont {B.}~\bibnamefont {Wardell}},\ }\href
  {\doibase 10.1103/PhysRevD.84.084001} {\bibfield  {journal} {\bibinfo
  {journal} {Phys.Rev.}\ }\textbf {\bibinfo {volume} {D84}},\ \bibinfo {pages}
  {084001} (\bibinfo {year} {2011})},\ \Eprint {http://arxiv.org/abs/1107.0012}
  {arXiv:1107.0012 [gr-qc]} \BibitemShut {NoStop}%
\bibitem [{\citenamefont {Detweiler}\ \emph {et~al.}(2003)\citenamefont
  {Detweiler}, \citenamefont {Messaritaki},\ and\ \citenamefont
  {Whiting}}]{Detweiler:2002gi}%
  \BibitemOpen
  \bibfield  {author} {\bibinfo {author} {\bibfnamefont {S.~L.}\ \bibnamefont
  {Detweiler}}, \bibinfo {author} {\bibfnamefont {E.}~\bibnamefont
  {Messaritaki}}, \ and\ \bibinfo {author} {\bibfnamefont {B.~F.}\ \bibnamefont
  {Whiting}},\ }\href {\doibase 10.1103/PhysRevD.67.104016} {\bibfield
  {journal} {\bibinfo  {journal} {Phys.Rev.}\ }\textbf {\bibinfo {volume}
  {D67}},\ \bibinfo {pages} {104016} (\bibinfo {year} {2003})},\ \Eprint
  {http://arxiv.org/abs/gr-qc/0205079} {arXiv:gr-qc/0205079 [gr-qc]}
  \BibitemShut {NoStop}%
\bibitem [{\citenamefont {Haas}\ and\ \citenamefont
  {Poisson}(2006)}]{Haas:2006ne}%
  \BibitemOpen
  \bibfield  {author} {\bibinfo {author} {\bibfnamefont {R.}~\bibnamefont
  {Haas}}\ and\ \bibinfo {author} {\bibfnamefont {E.}~\bibnamefont {Poisson}},\
  }\href {\doibase 10.1103/PhysRevD.74.044009} {\bibfield  {journal} {\bibinfo
  {journal} {Phys.Rev.}\ }\textbf {\bibinfo {volume} {D74}},\ \bibinfo {pages}
  {044009} (\bibinfo {year} {2006})},\ \Eprint
  {http://arxiv.org/abs/gr-qc/0605077} {arXiv:gr-qc/0605077 [gr-qc]}
  \BibitemShut {NoStop}%
\bibitem [{\citenamefont {Poisson}\ \emph {et~al.}(2011)\citenamefont
  {Poisson}, \citenamefont {Pound},\ and\ \citenamefont
  {Vega}}]{Poisson:2011nh}%
  \BibitemOpen
  \bibfield  {author} {\bibinfo {author} {\bibfnamefont {E.}~\bibnamefont
  {Poisson}}, \bibinfo {author} {\bibfnamefont {A.}~\bibnamefont {Pound}}, \
  and\ \bibinfo {author} {\bibfnamefont {I.}~\bibnamefont {Vega}},\ }\href@noop
  {} {\bibfield  {journal} {\bibinfo  {journal} {Living Rev.Rel.}\ }\textbf
  {\bibinfo {volume} {14}},\ \bibinfo {pages} {7} (\bibinfo {year} {2011})},\
  \Eprint {http://arxiv.org/abs/1102.0529} {arXiv:1102.0529 [gr-qc]}
  \BibitemShut {NoStop}%
\bibitem [{\citenamefont {Loffler}\ \emph {et~al.}(2012)\citenamefont
  {Loffler}, \citenamefont {Faber}, \citenamefont {Bentivegna}, \citenamefont
  {Bode}, \citenamefont {Diener} \emph {et~al.}}]{Loffler:2011ay}%
  \BibitemOpen
  \bibfield  {author} {\bibinfo {author} {\bibfnamefont {F.}~\bibnamefont
  {Loffler}}, \bibinfo {author} {\bibfnamefont {J.}~\bibnamefont {Faber}},
  \bibinfo {author} {\bibfnamefont {E.}~\bibnamefont {Bentivegna}}, \bibinfo
  {author} {\bibfnamefont {T.}~\bibnamefont {Bode}}, \bibinfo {author}
  {\bibfnamefont {P.}~\bibnamefont {Diener}},  \emph {et~al.},\ }\href
  {\doibase 10.1088/0264-9381/29/11/115001} {\bibfield  {journal} {\bibinfo
  {journal} {Class.Quant.Grav.}\ }\textbf {\bibinfo {volume} {29}},\ \bibinfo
  {pages} {115001} (\bibinfo {year} {2012})},\ \Eprint
  {http://arxiv.org/abs/1111.3344} {arXiv:1111.3344 [gr-qc]} \BibitemShut
  {NoStop}%
\bibitem [{\citenamefont {Goodale}\ \emph {et~al.}(2003)\citenamefont
  {Goodale}, \citenamefont {Allen}, \citenamefont {Lanfermann}, \citenamefont
  {Mass{\'o}}, \citenamefont {Radke}, \citenamefont {Seidel},\ and\
  \citenamefont {Shalf}}]{Goodale:2002a}%
  \BibitemOpen
  \bibfield  {author} {\bibinfo {author} {\bibfnamefont {T.}~\bibnamefont
  {Goodale}}, \bibinfo {author} {\bibfnamefont {G.}~\bibnamefont {Allen}},
  \bibinfo {author} {\bibfnamefont {G.}~\bibnamefont {Lanfermann}}, \bibinfo
  {author} {\bibfnamefont {J.}~\bibnamefont {Mass{\'o}}}, \bibinfo {author}
  {\bibfnamefont {T.}~\bibnamefont {Radke}}, \bibinfo {author} {\bibfnamefont
  {E.}~\bibnamefont {Seidel}}, \ and\ \bibinfo {author} {\bibfnamefont
  {J.}~\bibnamefont {Shalf}},\ }in\ \href {http://edoc.mpg.de/3341} {\emph
  {\bibinfo {booktitle} {Vector and Parallel Processing -- VECPAR'2002, 5th
  International Conference, Lecture Notes in Computer Science}}},\ Vol.\
  \bibinfo {volume} {2565}\ (\bibinfo  {publisher} {Springer},\ \bibinfo
  {address} {Berlin},\ \bibinfo {year} {2003})\BibitemShut {NoStop}%
\bibitem [{Cactus developers()}]{Cactuscode:web}%
  \BibitemOpen
  Cactus developers,\ \href {http://www.cactuscode.org/} {\enquote {\bibinfo
  {title} {{Cactus Computational Toolkit}},}\ }\bibinfo {note}
  {\url{http://www.cactuscode.org/}}\BibitemShut {NoStop}%
\bibitem [{\citenamefont {Schnetter}\ \emph {et~al.}(2004)\citenamefont
  {Schnetter}, \citenamefont {Hawley},\ and\ \citenamefont
  {Hawke}}]{Schnetter:2003rb}%
  \BibitemOpen
  \bibfield  {author} {\bibinfo {author} {\bibfnamefont {E.}~\bibnamefont
  {Schnetter}}, \bibinfo {author} {\bibfnamefont {S.~H.}\ \bibnamefont
  {Hawley}}, \ and\ \bibinfo {author} {\bibfnamefont {I.}~\bibnamefont
  {Hawke}},\ }\href {\doibase 10.1088/0264-9381/21/6/014} {\bibfield  {journal}
  {\bibinfo  {journal} {Class.Quant.Grav.}\ }\textbf {\bibinfo {volume} {21}},\
  \bibinfo {pages} {1465} (\bibinfo {year} {2004})},\ \Eprint
  {http://arxiv.org/abs/gr-qc/0310042} {arXiv:gr-qc/0310042 [gr-qc]}
  \BibitemShut {NoStop}%
\bibitem [{\citenamefont {Schnetter}()}]{CarpetCode:web}%
  \BibitemOpen
  \bibfield  {author} {\bibinfo {author} {\bibfnamefont {E.}~\bibnamefont
  {Schnetter}},\ }\href {http://www.carpetcode.org/} {}\bibinfo {note}
  {{Carpet}: Adaptive Mesh Refinement for the {Cactus} Framework}\BibitemShut
  {NoStop}%
\bibitem [{\citenamefont {Schnetter}\ \emph {et~al.}(2006)\citenamefont
  {Schnetter}, \citenamefont {Diener}, \citenamefont {Dorband},\ and\
  \citenamefont {Tiglio}}]{Schnetter:2006pg}%
  \BibitemOpen
  \bibfield  {author} {\bibinfo {author} {\bibfnamefont {E.}~\bibnamefont
  {Schnetter}}, \bibinfo {author} {\bibfnamefont {P.}~\bibnamefont {Diener}},
  \bibinfo {author} {\bibfnamefont {E.~N.}\ \bibnamefont {Dorband}}, \ and\
  \bibinfo {author} {\bibfnamefont {M.}~\bibnamefont {Tiglio}},\ }\href
  {\doibase 10.1088/0264-9381/23/16/S14} {\bibfield  {journal} {\bibinfo
  {journal} {Class.Quant.Grav.}\ }\textbf {\bibinfo {volume} {23}},\ \bibinfo
  {pages} {S553} (\bibinfo {year} {2006})},\ \Eprint
  {http://arxiv.org/abs/gr-qc/0602104} {arXiv:gr-qc/0602104 [gr-qc]}
  \BibitemShut {NoStop}%
\bibitem [{\citenamefont {Diener}\ \emph {et~al.}(2007)\citenamefont {Diener},
  \citenamefont {Dorband}, \citenamefont {Schnetter},\ and\ \citenamefont
  {Tiglio}}]{Diener:2005tn}%
  \BibitemOpen
  \bibfield  {author} {\bibinfo {author} {\bibfnamefont {P.}~\bibnamefont
  {Diener}}, \bibinfo {author} {\bibfnamefont {E.~N.}\ \bibnamefont {Dorband}},
  \bibinfo {author} {\bibfnamefont {E.}~\bibnamefont {Schnetter}}, \ and\
  \bibinfo {author} {\bibfnamefont {M.}~\bibnamefont {Tiglio}},\ }\href
  {\doibase 10.1007/s10915-006-9123-7} {\bibfield  {journal} {\bibinfo
  {journal} {J.Sci.Comput.}\ }\textbf {\bibinfo {volume} {32}},\ \bibinfo
  {pages} {109} (\bibinfo {year} {2007})},\ \Eprint
  {http://arxiv.org/abs/gr-qc/0512001} {arXiv:gr-qc/0512001 [gr-qc]}
  \BibitemShut {NoStop}%
\bibitem [{\citenamefont {Dorband}\ \emph {et~al.}(2006)\citenamefont
  {Dorband}, \citenamefont {Berti}, \citenamefont {Diener}, \citenamefont
  {Schnetter},\ and\ \citenamefont {Tiglio}}]{Dorband:2006gg}%
  \BibitemOpen
  \bibfield  {author} {\bibinfo {author} {\bibfnamefont {E.~N.}\ \bibnamefont
  {Dorband}}, \bibinfo {author} {\bibfnamefont {E.}~\bibnamefont {Berti}},
  \bibinfo {author} {\bibfnamefont {P.}~\bibnamefont {Diener}}, \bibinfo
  {author} {\bibfnamefont {E.}~\bibnamefont {Schnetter}}, \ and\ \bibinfo
  {author} {\bibfnamefont {M.}~\bibnamefont {Tiglio}},\ }\href {\doibase
  10.1103/PhysRevD.74.084028} {\bibfield  {journal} {\bibinfo  {journal}
  {Phys.Rev.}\ }\textbf {\bibinfo {volume} {D74}},\ \bibinfo {pages} {084028}
  (\bibinfo {year} {2006})},\ \Eprint {http://arxiv.org/abs/gr-qc/0608091}
  {arXiv:gr-qc/0608091 [gr-qc]} \BibitemShut {NoStop}%
\end{thebibliography}%

\end{document}